\documentclass[onecollarge,final,amsfonts,amsmath,amssymb,aps,indentfirst,latexsym,mathrsfs,preprint,preprintnumbers,showpacks]{svjour2}

\usepackage{amsfonts,amsmath,amssymb,indentfirst,latexsym,mathrsfs}
\usepackage[mathscr]{eucal}
\usepackage[english]{babel}
\usepackage[T1]{fontenc}
\usepackage[cp1250]{inputenc}

\def\beq{\begin{equation}}
\def\eeq{\end{equation}}
\def\beqs{\begin{equation*}}
\def\eeqs{\end{equation*}}
\def\beqa{\begin{eqnarray}}
\def\eeqa{\end{eqnarray}}
\def\beqas{\begin{eqnarray*}}
\def\eeqas{\end{eqnarray*}}

\begin{document}

\title{Constraints and Interactions in Quantization\\ of Yukawa Model with Higher Order Derivatives} 

\author{Jan {\.Z}ochowski}
\institute{Faculty of Physics, University of Bia{\l}ystok, ul. Cio{\l}kowskiego 1L, 15-245 Bia{\l}ystok, Poland\\
\email{j.zochowski@uwb.edu.pl}}

\maketitle

%\PACS{}

\begin{abstract}

This work is dedicated to the quantization of the light-front Yukawa model in $D=1+3$ dimensions with higher order derivatives of the scalar field. The problem of the computing Dirac brackets and the (anti-) commutator algebra of interacting fields in the presence of the constraints is discussed. The Dirac method and the Ostrogradski formalism of the higher order derivatives are exploited. The systematic method of obtaining the inverse of the functional Dirac-Bergmann matrix with interactions and higher order derivatives is introduced in two variants. The discussion of applications and details of these two variants are conducted. The results of the quantization in the form of the (anti-) commutator algebra are presented and analyzed with special regard to the structure of the interactions for the light-front Yukawa model, which includes the higher order derivatives.\\


\end{abstract}

\section{Yukawa Model with Higher Order Derivatives and Its Canonical Structure}
\label{sec1}

It is well known, the main problem of the canonical formalism is to introduce the relevant quantum fields in order to obtain the acceptable Hamiltonian density from the Lagrangian one, which should be entirely agreeable with the constraints including the assumptive interactions. The natural structure, which indicates the adequate quantum field operators for the Hamiltonian density is the algebra of the (anti-) commutators. This ought to be derived via the method, introduced by Dirac, and leading from the canonical Poisson brackets to just so Dirac ones, 
consistent with the constraints and being subjected to the procedure of the quantization at the last step. In the vast majority of cases the constraints include complete structure of the interactions, since they are, among others, but more often than not, the equations of motion for the non-dynamical degrees of freedom. Finally, if we are able, despite mathematical difficulties, to incorporate the constraints with interactions into the Dirac procedure, the detailed solution of this allows us to obtain the commutative relationships of the interacting fields. And furthermore, the problem of relevant selection of the dynamical and the non-dynamical fields for the canonical formalism is especially interesting and worth of taking trouble in the case of the models with higher order derivatives. One of the simplest version of the theories belonging to mentioned class is the Yukawa model in $D=1+3$ dimensions, but extended by the minimal, just including the higher order derivative term, introduced in the below Lagrangian density. Let's say for order, that except this contribution our model embraces one massive flavour of the fermionic and one of the massive scalar field (self-coupled) and the typical Yukawa coupling term 
\beq
\label{ldym}
{\cal L}=
{\overline\Psi}\left(i\!\!\not\!\partial-M\right)\Psi+
{1\over 2}\left(\partial\phi\right)^{2}-
{1\over 2}m^{2}\phi^{2}-
{1\over 4}\lambda\phi^{4}-
g\phi{\overline\Psi}\Psi+
{1\over 2}\alpha\left(\partial^{2}\phi\right)^{2}. 
\eeq
The parameter $\lambda$ denotes in this formula the dimensionless coupling constant for the self-interaction of the scalar field. Also dimensionless $g$ marks the Yukawa coupling constant at the tri-linear vertex between the scalar field $\phi$ of the mass $m$, the incoming fermionic field $\Psi$ and the outgoing one ${\overline\Psi}$, both of the unique flavour of the mass $M$. Remarkably, the constant $\alpha$, multiplying the higher order derivative term, has the dimension equal to $-2$. This term makes up, in fact, the essentially simplest version \cite{b1,b2,b3} of the three-parameter Bernard-Duncan model for the single scalar and real field 
\beq
\label{berdun}
{\cal L}_{\!\!\!\mbox{ }_{B-D}}= 
{1\over 2}\alpha\left(\partial^{2}\phi\right)^{2}-
{1\over 2}\varepsilon\left(\partial\phi\right)^{2}+
{1\over 2}\eta\phi^{2}. 
\eeq
Herein, two of these parameters, $\varepsilon$ and $\eta$, are put, to the needs of our analysis, as equal to zero. Only $\alpha$, the important one, remains non-trivial. It should be honestly emphasized at this point, the implementation of the higher order term into the Yukawa model and other theories leads to many more or less basic problems, like the energy ladder of the states unconstrained from below and others \cite{b2}. Nevertheless, the Yukawa model with higher order derivative term (\ref{ldym}) stands as a handy area for testing the Dirac formalism of quantization with the constraints, which also embrace the interactions. 

The structure of the canonical momenta for our model (\ref{ldym}) seems to be simpler for the light-front formulation than for the equal time one \cite{b4}, what tempts us to write down this Lagrangian density in that's coordinates: (\ref{lfcoord}), (\ref{lfhypcoord}), where $x^{+}$ is the variable of evolution. The light-front Yukawa model with higher order derivatives should now be rewritten, for convenience, on the grounds of the fermionic bispinor notation: (\ref{spindecomppsi}), (\ref{spinors}) and (\ref{uconstpmspin}). As a consequence, we have 
\beq
\label{ldymlf}
{\cal L}=
\eeq
\beqs
=\!i\sqrt{2}\;\Psi^{\dagger}_{+}\partial_{+}\Psi_{+}+
i\sqrt{2}\;\Psi^{\dagger}_{-}\partial_{-}\Psi_{-}+
{i\over{\sqrt{2}}}
\Psi^{\dagger}_{+}\gamma^{-}\gamma^{j}\partial_{j}\Psi_{-}+
{i\over{\sqrt{2}}}
\Psi^{\dagger}_{-}\gamma^{+}\gamma^{j}\partial_{j}\Psi_{+}-
{1\over{\sqrt{2}}}M
\Psi^{\dagger}_{+}\gamma^{-}\Psi_{-}-
{1\over{\sqrt{2}}}M
\Psi^{\dagger}_{-}\gamma^{+}\Psi_{+}+
\eeqs
\beqs
+\left(\partial_{+}\phi\right)\left(\partial_{-}\phi\right)-
{1\over 2}\left(\partial_{j}\phi\right)\left(\partial_{j}\phi\right)-
{1\over 2}m^{2}\phi^{2}-
{1\over 4}\lambda\phi^{4}-
{1\over{\sqrt{2}}}g
\phi\;\Psi^{\dagger}_{+}\gamma^{-}\Psi_{-}-
{1\over{\sqrt{2}}}g
\phi\;\Psi^{\dagger}_{-}\gamma^{+}\Psi_{+}+
\eeqs
\beqs
+{1\over 2}\alpha
\left(\partial_{+}\partial_{-}\phi+
\partial_{-}\partial_{+}\phi-
\partial_{j}\partial_{j}\phi\right)^{2}.
\eeqs
From point of view of the classification of the Lagrangians, analyzed model includes the fermionic sector composed of the fundamental fields $\Psi_{\pm}$, their conjugation $\Psi^{\dagger}_{\pm}$ and their first order derivatives. Thus, the principle of minimal action requires in this case only the standard approach and leads to the ordinary Lagrange equations: 
\beq
\label{ordlagr}
{{\partial{\cal L}}\over{\partial{\Psi_{\pm}}}}-
\partial_{\mu}\left({{\partial{\cal L}}\over{\partial\left({\partial_{\mu}\Psi_{\pm}}\right)}}\right)
=0, \;\;\;\;\;\;\;\;\;\;\;\; 
{{\partial{\cal L}}\over{\partial{\Psi^{\dagger}_{\pm}}}}-
\partial_{\mu}\left({{\partial{\cal L}}\over{\partial\left({\partial_{\mu}\Psi^{\dagger}_{\pm}}\right)}}\right)
=0,  
\eeq
wherein the summation index runs $\mu=+,-,j$ and $j=1,2$. We also have the second order derivatives for the scalar field within the Lagrangian density, what obliges us to apply the Ostrogradski formalism in this sector. So, in other words, the scalar sector of our model embraces the $\phi$ and its first order derivative by the variable of evolution $\partial_{+}\phi$, as the independent fields. According to the Ostrogradski approach \cite{b5,b6,b7} for this case of the Lagrangian density, which includes the derivatives up to the $n$th order of the single scalar field ${\cal L}(\phi,\phi_{,\mu_{1}},\dots,\phi_{,\mu_{1}\dots\mu_{n}})$, where $\phi_{,\mu_{1}\dots\mu_{r}}\equiv\partial_{\mu_{1}}\dots\partial_{\mu_{r}}\phi$ and $r=1,\dots,n$, the relevant equation of motion is 
\beq
\label{hiordlagr}
{{\partial{\cal L}}\over{\partial{\phi}}}-
\partial_{\mu_{1}}\left({{\partial{\cal L}}\over{\partial\phi_{,\mu_{1}}}}\right)
+\dots+
(-1)^{n}\partial_{\mu_{1}}\dots\partial_{\mu_{n}}
\left({{\partial{\cal L}}\over{\partial\phi_{,\mu_{1}\dots\mu_{n}}}}\right)
=0.
\eeq
We have $n=2$ in our case. Thus, the equations of motion for the Yukawa model with higher order derivatives, inferred from the formulae (\ref{ordlagr}) and (\ref{hiordlagr}), yield respectively: 
\beq
\label{empsim}
2i\partial_{+}\Psi_{+}+
i\gamma^{-}\!\!\not\!{\stackrel{\!\!\rightarrow}{\partial_{\perp}}}\Psi_{-}-
\left(M+g\phi\right)\gamma^{-}\Psi_{-}
=0, \;\;\;\;\;\; 
2i\partial_{+}\Psi^{\dagger}_{+}-
i\Psi^{\dagger}_{-}\!\!\not\!{\stackrel{\!\!\leftarrow}{\partial_{\perp}}}\gamma^{+}+
\left(M+g\phi\right)\Psi^{\dagger}_{-}\gamma^{+}
=0,
\eeq
\beq
\label{empsim1}
2i\partial_{-}\Psi_{-}+
i\gamma^{+}\!\!\not\!{\stackrel{\!\!\rightarrow}{\partial_{\perp}}}\Psi_{+}-
\left(M+g\phi\right)\gamma^{+}\Psi_{+}
=0, \;\;\;\;\;\; 
2i\partial_{-}\Psi^{\dagger}_{-}-
i\Psi^{\dagger}_{+}\!\!\not\!{\stackrel{\!\!\leftarrow}{\partial_{\perp}}}\gamma^{-}+
\left(M+g\phi\right)\Psi^{\dagger}_{+}\gamma^{-}
=0,
\eeq
\beq
\label{emphi}
\left(\partial_{+}\partial_{-}\phi+\partial_{-}\partial_{+}\phi-
\partial_{j}\partial_{j}\phi\right)+
m^{2}\phi+\lambda\phi^{3}+
{1\over{\sqrt{2}}}g
\left(\Psi^{\dagger}_{+}\gamma^{-}\Psi_{-}+
\Psi^{\dagger}_{-}\gamma^{+}\Psi_{+}\right)-
\eeq
\beqs
-\alpha
\left(\partial_{+}\partial_{-}+\partial_{-}\partial_{+}-
\partial_{k}\partial_{k}\right)
\left(\partial_{+}\partial_{-}\phi+\partial_{-}\partial_{+}\phi-
\partial_{l}\partial_{l}\phi\right)
=0,
\eeqs
where the arrows are related to the direction of the derivative action. 

The canonical structure of the Poisson brackets is determined in details by the precise choice of the momenta. This is especially important for the theories with higher order derivatives, like just discussed Yukawa model. In a very general approach, more mathematical, all partial derivatives with respect to all partial derivatives of the fields, regardless they are dynamical or not, {\it i.e.} they are physical degrees of freedom or not, may be treated the same manner, as the contributions to the Legendre transformation. However, this formalism sometimes obliterates the physical distinction between dynamical and non-dynamical degrees of freedom, especially for the theories with the flat space-time background - in absence of the gravity \cite{b8,b9}. Hence, we decide to maintain the division into dynamical and non-dynamical momenta. Only these first belong to the canonical structure of our model (see \cite{b2} versus \cite{b6}). Thus, we have in our model four standard canonical momenta, conjugated with respect to the relevant fermionic fields: 
\beq
\label{defcanmom}
\pi_{\pm}=
{{\partial{\cal L}}\over{\partial(\partial_{+}\Psi^{\dagger}_{\pm})}}, 
\;\;\;\;\;\;\;\;\;\;\;\;\;\;\;\;\;\;     
\pi^{\dagger}_{\pm}=
{{\partial{\cal L}}\over{\partial(\partial_{+}\Psi_{\pm})}}. 
\eeq 
They can be obtained from the Lagrange formalism without the Ostrogradski approach to the higher order derivatives, which are not present in this sector. When taking into account, that our Lagrangian density has the form (\ref{ldymlf}), we may write: 
\beq
\label{cmphipsippsim}
\pi_{+}=0, \;\;\;\;\;\;\;\;\;
\pi^{\dagger}_{+}=
i\sqrt{2}\;\Psi^{\dagger}_{+}, \;\;\;\;\;\;\;\;\;
\pi_{-}=0, \;\;\;\;\;\;\;\;\;  
\pi^{\dagger}_{-}=0.
\eeq 
The scalar sector of our model is just that, which demands aforementioned Ostrogradski method for the treatment of the higher order derivatives. This formalism requires to introduce the $n$ canonical momenta, if the derivatives of the field within the Lagrangian density are from the first up to the $n$th order. They are canonically conjugated with relevant derivatives of the scalar field: 
\beq
\label{defcanhiordmom} 
\pi^{\mu_{1}\dots\mu_{n}}_{\phi}=
{{{\partial{\cal L}}}\over{\partial\phi_{,\mu_{1}\dots\mu_{n}}}} , \;\;\;\;\;\;  
\pi^{\mu_{1}\dots\mu_{r}}_{\phi}=
{{{\partial{\cal L}}}\over{\partial\phi_{,\mu_{1}\dots\mu_{r}}}} -
\partial_{\mu_{r+1}}\pi^{\mu_{1}\dots\mu_{r}\mu_{r+1}}_{\phi}, \;\;\;\;\;\; 
r=1,\dots,n-1. 
\eeq
As indicated previously, analyzed Yukawa model includes, besides the $\phi$, also the first order derivative $\partial_{+}\phi$, as the independent field. The variable of evolution for this model in the light-front formulation is $x^{+}$, what makes, according to committed remarks above, that with help of the definition (\ref{defcanhiordmom}) we should introduce two dynamical momenta for the scalar sector, conjugated with the fields $\partial_{+}\phi$ and $\phi$, respectively: 
\beq
\label{canhiordscalmom}
\pi^{++}_{\phi}=0, \;\;\;\;\;\;\;\;\;\;\;\; 
\pi^{+}_{\phi}=
\partial_{-}\phi-
\alpha\left(
\partial_{-}\partial_{+}\partial_{-}\phi+
\partial_{-}\partial_{-}\partial_{+}\phi-
\partial_{-}\partial_{j}\partial_{j}\phi
\right).
\eeq     
In contrast, rest of the momenta: $\pi^{+-}_{\phi}$, $\pi^{+j}_{\phi}$, $\pi^{-+}_{\phi}$, $\pi^{--}_{\phi}$, $\pi^{-j}_{\phi}$, $\pi^{j+}_{\phi}$, $\pi^{j-}_{\phi}$, $\pi^{jk}_{\phi}$, $\pi^{-}_{\phi}$ and $\pi^{j}_{\phi}$, where $j,k=1,2$, are the non-dynamical ones. They are not the part of the canonical structure of discussed model and hence, the derivatives: $\partial_{-}\phi$, $\partial_{j}\phi$ cannot be treated as the independent fields in the approach presented here. 

We have at our disposal already introduced set of the dynamical momenta, canonically conjugated with the fundamental fields and then, we can postulate the algebra of the Poisson brackets onto the light-front hyper-surface: 
\beq
\label{canquant1}
\left\{
\Psi_{\pm}\left(x^{+},{\bar x}\right) \; , \; \pi^{\dagger}_{\pm}\left(x^{+},{\bar y}\right)
\right\}_{\!\!P}\!\!\!=
\delta^{(3)}\!\left({\bar x}-{\bar y}\right)\Lambda_{\pm}, \;\;\;\;\;\;
\left\{
\Psi^{\dagger}_{\pm}\left(x^{+},{\bar x}\right) \; , \; \pi_{\pm}\left(x^{+},{\bar y}\right) 
\right\}_{\!\!P}\!\!\!=
\delta^{(3)}\!\left({\bar x}-{\bar y}\right)\Lambda_{\pm},
\eeq
\beq
\label{canquant3}
\left\{
\partial_{+}\phi\left(x^{+},{\bar x}\right) \; , \; \pi^{++}_{\phi}\left(x^{+},{\bar y}\right)
\right\}_{\!\!P}=
\delta^{(3)}\left({\bar x}-{\bar y}\right), \;\;\;\;\;\; 
\left\{
\phi\left(x^{+},{\bar x}\right) \; , \; \pi^{+}_{\phi}\left(x^{+},{\bar y}\right)
\right\}_{\!\!P}=
\delta^{(3)}\left({\bar x}-{\bar y}\right).
\eeq
Let's observe, there is the tensor product of the spinor fields within the fermionic Poisson brackets and due to this fact, the projectors $\Lambda_{\pm}$, defined by the expressions (\ref{tenprodlambdapm}) and (\ref{psipmaslambdapsi}), appear in relevant patterns above. These canonical Poisson brackets are solely non-vanishing elements of their algebra in the case of the light-front Yukawa model with higher order derivatives, discussed in this work. For the reasons mentioned in all points of the discussion, this algebra does not embrace the brackets for the fields $\partial_{-}\phi$, $\partial_{j}\phi$ and for the momenta, conjugated with them. 

\section{Constraints of Yukawa Model with Higher Order Derivatives}
\label{dbphri}

Analyzed system, the Yukawa model, is described by the Lagrangian density (\ref{ldymlf}). This embraces the dynamical equations of motion (\ref{empsim}) and (\ref{emphi}), which are determining the $x^{+}$ evolution of the independent fields $\Psi_{+}$ and $\phi$, respectively. Thus, the equation $\pi^{\dagger}_{+}=i\sqrt{2}\Psi^{\dagger}_{+}$ defines relevant momentum to the needs of the first formula within the algebra (\ref{canquant1}). Of course, these equations: (\ref{empsim}), (\ref{emphi}) and the above pattern for $\pi^{\dagger}_{+}$ are not the constraints. 

However, comparison of the formulae (\ref{canquant1}) and (\ref{canquant3}) on the one hand to the first, to the third, to the fourth of the (\ref{cmphipsippsim}) and to the relationship (\ref{canhiordscalmom}) on the other, discloses that they are mutually inconsistent. We should use the Dirac procedure for the equations of the momenta: $\pi^{\dagger}_{-}=0$, $\pi_{-}=0$, $\pi^{++}_{\phi}=0$,  $\pi^{+}_{\phi}=\partial_{-}\phi-\alpha\partial_{-}\partial^{2}\phi$ and $\pi_{+}=0$ to identify the constraints, which emerging in our model. We cannot express the derivative of the fundamental field with respect to the variable of evolution by this field and by the relevant canonically conjugated momentum \cite{b10,b11,b12} in the case of the primary constraints. The necessary and the sufficient condition for appearance of the primary constraint is to have the singular matrix of the second derivative of the Lagrangian density with respect to the fundamental derivatives of the field by the variable of evolution \cite{b13,b14}. As can be seen, the light-front Yukawa model with higher order derivatives exhibits the above property. The Lagrangian density (\ref{ldymlf}) is, so called, irregular one in the fermionic sector, wherein we have $\partial{\cal L}/\partial(\partial_{+}\Psi_{-})=0$, what leads to the conclusion 
$\det[\partial^{2}{\cal L}/\partial(\partial_{+}\Psi_{a})\partial(\partial_{+}\Psi_{b})]=0$ with $a,b=+,-$. The same case we have for the fundamental field $\Psi^{\dagger}_{-}$. We may write down, in the context of these remarks, two equations of the primary constraints of analyzed model $\Phi_{1,2}$, denoted below with help of the symbol $\approx$. They are: 
\beq
\label{constraintsprimary}
\Phi_{1}\approx\pi^{\dagger}_{-}=0, 
\;\;\;\;\;\;\;\;\;\;\;\;\;\;\;  
\Phi_{2}\approx\pi_{-}=0.
\eeq
We come across, by the study on the scalar sector of the Yukawa model, completely the same situation. The irregular Lagrangian density (\ref{ldymlf}) gives $\partial{\cal L}/\partial(\partial_{+}\partial_{+}\phi)=0$ and then consequently infers, that 
$\det[\partial^{2}{\cal L}/\partial(\partial_{+}\partial_{\mu}\phi)\partial(\partial_{+}\partial_{\nu}\phi)]=0$, where $\mu,\nu=+,-,j$ and $j=1,2$. Therefore, we can put another primary constraint 
\beq
\label{constraintprimary0}
\Phi_{3}\approx\pi^{++}_{\phi}=0.
\eeq
We find, by the same type of analysis, the fourth primary constraint, related to the momentum $\pi^{+}_{\phi}$ and introduced by the second pattern (\ref{canhiordscalmom}). We are simply convincing, that happens the condition    
$\det[\partial\pi^{+}_{\phi}/\partial(\partial_{+}\phi)]=\partial\pi^{+}_{\phi}/\partial(\partial_{+}\phi)=0$, what complies with the necessary and the sufficient condition for the primary constraint. As a result 
\beq
\label{constraintsprimary1}
\Phi_{4}\approx
\pi^{+}_{\phi}-\partial_{-}\phi+
\alpha\left(
\partial_{-}\partial_{+}\partial_{-}\phi+
\partial_{-}\partial_{-}\partial_{+}\phi-
\partial_{-}\partial_{j}\partial_{j}\phi
\right)=0.
\eeq
Let's remark, the specific case provides the dynamical momentum $\pi_{+}=0$, conjugated with the field $\Psi^{\dagger}_{+}$. This manifestly belongs to the set of the primary constraints within analyzed model. Nevertheless, a special treatment is allowed for this constraint. One can clearly see, that each of the equations (\ref{empsim}) may be obtained from another by the Hermitian conjugation. Thus, the dynamic and the canonical structure for the field $\Psi^{\dagger}_{+}$, conjugated with respect to the $\pi_{+}=0$, may be derived from those for the $\Psi_{+}$. We can eliminate in consequence the $\Psi^{\dagger}_{+}$ from the considerations and reduce by one the set of the constraints, skipping $\pi_{+}=0$. This saturates the discussion on the primary constraints in our case. The Yukawa model with higher order derivatives also embraces the secondary constraints. They emerge as the additional equations for these of the fields, which of them are conjugated with the momenta involved in the primary constraints \cite{b13,b15,b16}. Herein, such the conditions of the consistency for the primary constraints are the relevant non-dynamical Lagrange equations: 
\beq
\label{constraintssecondary4} 
\Phi_{5}\approx
2i\partial_{-}\Psi^{\dagger}_{-}-
i\Psi^{\dagger}_{+}\!\not\!\!{\stackrel{\leftarrow}{\partial}}_{\!\perp}\gamma^{-}+
\left(M+g\phi\right)\Psi^{\dagger}_{+}\gamma^{-}
=0,
\eeq
\beq
\label{constraintssecondary5}
\Phi_{6}\approx
2i\partial_{-}\Psi_{-}+
i\gamma^{+}\!\!\not\!\!{\stackrel{\rightarrow}{\partial}}_{\!\perp}\Psi_{+}-
\left(M+g\phi\right)\gamma^{+}\Psi_{+}
=0. 
\eeq 
It should be made two comments at this point. Firstly, we consider both, the primary and the secondary, constraints together, during  implementation of the Dirac procedure. Secondly, we obtain the Dirac brackets and then the (anti-) commutators by incorporation of the complete structure of the interactions. Let's introduce the Dirac-Bergmann matrix, which elements should be defined as the Poisson brackets of the constraints. Within realities of the light-front coordinates, relevant for the formulation of our model, this satisfies 
\beq
\label{fmatrix}
{\cal F}_{kl}\left(x^{+},{\bar w},{\bar z}\right)=
\left\{\Phi_{k}\left(x^{+},{\bar w}\right) \; , \; \Phi_{l}\left(x^{+},{\bar z}\right)\right\}_{\!\!P}.
\eeq
The constraints $\Phi_{1,\dots,6}$ are determined by the patterns: (\ref{constraintsprimary}), (\ref{constraintprimary0}), (\ref{constraintsprimary1}), (\ref{constraintssecondary4}) and (\ref{constraintssecondary5}), for the case studied in this work. It is easy to convince, that the Poisson brackets for none of the $\Phi_{k}$, where $k=1,\dots,6$, are not zeroed with all other constraints. For instance, we have the non-trivial Poisson brackets: $\{\Phi_{1},\Phi_{6}\}_{\!\!\!\!\mbox{ }_{P}}\neq0$, $\{\Phi_{2},\Phi_{5}\}_{\!\!\!\!\mbox{ }_{P}}\neq0$, $\{\Phi_{3},\Phi_{4}\}_{\!\!\!\!\mbox{ }_{P}}\neq0$, $\{\Phi_{5},\Phi_{6}\}_{\!\!\!\!\mbox{ }_{P}}\neq0$ and so forth. Noticeably, the eliminated constraint $\pi_{+}=0$ also belongs to these of the second class, due to $\{\pi_{+},\Phi_{5}\}_{\!\!\!\!\mbox{ }_{P}}\neq0$. Thus, our system does not embrace the constraints of the first class. Finally, we ensure, that the Dirac-Bergmann matrix for the Yukawa model with higher order derivatives is non-singular. Therefore, according to the definition, the complete set of our constraints consists of only these of the second class. This means, that our constraints may be directly used for calculations of the Dirac brackets and then the (anti-) commutators. The matrix of the Poisson brackets of the constraints of the second class ${\cal F}(x^{+},{\bar w},{\bar z})$ should be derived with help of the canonical quantization rules (\ref{canquant1}) and (\ref{canquant3}). We take them for the needs of current discussion and we obtain precise structure of the ${\cal F}(x^{+},{\bar w},{\bar z})$ array  
\beq
\label{fmatrixelements}
{\cal F}\left(x^{+},{\bar w},{\bar z}\right)=
\eeq
\beqs
=\left(\;
\begin{array}{cccccc}
0&\;\;0&\;\;0&\;\;0&\;\;0&\;\;{\cal F}_{16}\left(x^{+},{\bar w},{\bar z}\right)\\
0&\;\;0&\;\;0&\;\;0&\;\;{\cal F}_{25}\left(x^{+},{\bar w},{\bar z}\right)&\;\;0\\
0&\;\;0&\;\;0&\;\;{\cal F}_{34}\left(x^{+},{\bar w},{\bar z}\right)&\;\;0&\;\;0\\
0&\;\;0&\;\;
{\cal F}_{43}\left(x^{+},{\bar w},{\bar z}\right)&\;\;
{\cal F}_{44}\left(x^{+},{\bar w},{\bar z}\right)&\;\;
{\cal F}_{45}\left(x^{+},{\bar w},{\bar z}\right)&\;\;
{\cal F}_{46}\left(x^{+},{\bar w},{\bar z}\right)\\
0&
\;\;{\cal F}_{52}\left(x^{+},{\bar w},{\bar z}\right)&\;\;
0&\;\;
{\cal F}_{54}\left(x^{+},{\bar w},{\bar z}\right)&\;\;
0&\;\;
{\cal F}_{56}\left(x^{+},{\bar w},{\bar z}\right)\\ 
{\cal F}_{61}\left(x^{+},{\bar w},{\bar z}\right)&\;\;
0&\;\;0&\;\;
{\cal F}_{64}\left(x^{+},{\bar w},{\bar z}\right)&\;\;
{\cal F}_{65}\left(x^{+},{\bar w},{\bar z}\right)&\;\; 
0\\
\end{array}
\right). 
\eeqs
Herein, we have: 
\beq
\label{fmatrixelements1}
{\cal F}_{16}\left(x^{+},{\bar w},{\bar z}\right)=
2i\partial^{w}_{-}
\delta^{(3)}\left({\bar w}-{\bar z}\right)\Lambda_{-}, \;\;\;\;\;  
{\cal F}_{25}\left(x^{+},{\bar w},{\bar z}\right)=
2i\partial^{w}_{-}
\delta^{(3)}\left({\bar w}-{\bar z}\right)\Lambda_{-}, 
\eeq
\beq
\label{fmatrixelements2}
{\cal F}_{34}\left(x^{+},{\bar w},{\bar z}\right)=
-2\alpha\partial^{2w}_{-}
\delta^{(3)}\left({\bar w}-{\bar z}\right){\rm P}, \;\;\;\;\;\; 
{\cal F}_{43}\left(x^{+},{\bar w},{\bar z}\right)=
2\alpha\partial^{2w}_{-}
\delta^{(3)}\left({\bar w}-{\bar z}\right){\rm P}, 
\eeq
\beq
\label{fmatrixelements31}
{\cal F}_{44}\left(x^{+},{\bar w},{\bar z}\right)=
-2\partial^{w}_{-}
\delta^{(3)}\left({\bar w}-{\bar z}\right){\rm P}-
2\alpha\partial^{w}_{-}\Delta^{w}_{\perp}
\delta^{(3)}\left({\bar w}-{\bar z}\right){\rm P},  
\eeq
\beq
\label{fmatrixelements32}
{\cal F}_{45}\left(x^{+},{\bar w},{\bar z}\right)=
-g\Psi^{\dagger}_{+}\left(x^{+},{\bar z}\right)\gamma^{-}
\delta^{(3)}\left({\bar w}-{\bar z}\right), 
\eeq
\beq
\label{fmatrixelements4}
{\cal F}_{46}\left(x^{+},{\bar w},{\bar z}\right)=
g\gamma^{+}\Psi_{+}\left(x^{+},{\bar z}\right)
\delta^{(3)}\left({\bar w}-{\bar z}\right), \;\;\;\;\;\;   
{\cal F}_{52}\left(x^{+},{\bar w},{\bar z}\right)=
-2i\partial^{w}_{-}
\delta^{(3)}\left({\bar w}-{\bar z}\right)\Lambda_{-}, 
\eeq 
\beq
\label{fmatrixelements5}
{\cal F}_{54}\left(x^{+},{\bar w},{\bar z}\right)=
g\Psi^{\dagger}_{+}\left(x^{+},{\bar w}\right)\gamma^{-}
\delta^{(3)}\left({\bar w}-{\bar z}\right), \;\;\;\;\;\;   
{\cal F}_{56}\left(x^{+},{\bar w},{\bar z}\right)=
{\xi}\left(x^{+},{\bar w},{\bar z}\right)
\delta^{(3)}\left({\bar w}-{\bar z}\right),
\eeq
\beq
\label{fmatrixelements6}   
{\cal F}_{61}\left(x^{+},{\bar w},{\bar z}\right)=
-2i\partial^{w}_{-}
\delta^{(3)}\left({\bar w}-{\bar z}\right)\Lambda_{-}, \;\;\;\;\;\;  
{\cal F}_{64}\left(x^{+},{\bar w},{\bar z}\right)=
-g\gamma^{+}\Psi_{+}\left(x^{+},{\bar w}\right)
\delta^{(3)}\left({\bar w}-{\bar z}\right), 
\eeq
\beq
\label{fmatrixelements7}  
{\cal F}_{65}\left(x^{+},{\bar w},{\bar z}\right)=
-{\xi}\left(x^{+},{\bar w},{\bar z}\right)
\delta^{(3)}\left({\bar w}-{\bar z}\right).  
\eeq
The differential operator $\xi(x^{+},{\bar w},{\bar z})$, acting on the Dirac delta, is determined by the expression  
\beq
\label{amatrixfunction}
{\xi}\left(x^{+},{\bar w},{\bar z}\right)=
i\sqrt{2}\Delta^{w}_{\perp}\Lambda_{-}-
i\sqrt{2}\left[M+g\phi\left(x^{+},{\bar w}\right)\right]\!
\left[M+g\phi\left(x^{+},{\bar z}\right)\right]
\Lambda_{-}. 
\eeq 
The symbol ${\rm P}$ denotes the projection operator. Precise form of this ${\rm P}$ could be inferred from the equation (\ref{dirbracketdef}). According to this pattern, the block structure of the ${\cal F}^{-1}$ must be matched to the matrix of the Poisson brackets. They are given by the tensor product of their arguments $\psi_{a}$ and $\psi_{b}$. Therefore, we can have 
${\rm P}={\rm I},\Lambda_{\pm},\gamma^{\pm},\gamma^{\mu}\otimes\gamma^{\nu}$ and $\mu,\nu=+,-$ in the case of the Yukawa model with higher order derivatives, due to the decompositions (\ref{spindecomppsi}) - (\ref{gammapsi}) of the bispinors. The size of the unit matrix ${\rm I}$ is $2\times2$. 

\section{Dirac Procedure for Yukawa Model with Higher Order Derivatives}
\label{dbphri1}

Dirac introduced the brackets being consistent with the constraints \cite{b17,b18,b19} and showed, that the pass through from the Poisson brackets to those first ought to be described by the prescription, which includes the inverse of the matrix (\ref{fmatrix}). In terms of the light-front coordinates, this prescription yields 
\beq
\label{dirbracketdef}
\left\{\psi_{a}\left(x^{+},{\bar x}\right) \; , \; \psi_{b}\left(x^{+},{\bar y}\right)\right\}_{\!\!D}=
\left\{\psi_{a}\left(x^{+},{\bar x}\right) \; , \; \psi_{b}\left(x^{+},{\bar y}\right)\right\}_{\!\!P}-
\eeq
\beqs
-\sum_{k,l=1}^{N}
\int_{R^{3}}\!{d^{3}{\bar w}}\int_{R^{3}}\!{d^{3}{\bar z}}\;
\left\{\psi_{a}\left(x^{+},{\bar x}\right) \; , \; \Phi_{k}\left(x^{+},{\bar w}\right)\right\}_{\!\!P}
{\cal F}^{-1}_{kl}\left(x^{+},{\bar w},{\bar z}\right)
\left\{\Phi_{l}\left(x^{+},{\bar z}\right) \; , \; \psi_{b}\left(x^{+},{\bar y}\right)\right\}_{\!\!P},
\eeqs  
where $\psi_{a}(x^{+},{\bar x})$ and $\psi_{b}(x^{+},{\bar y})$ denote, with their possible composite indexes $a$ and $b$, the fields apparent in the model. The parameter $N$ is the number of the second class constraints herein, showing also the size of the matrix (\ref{fmatrix}) and its inverse. We have $\psi_{a},\psi_{b}=\Psi_{+},\Psi_{-},\Psi^{\dagger}_{-},\phi,\partial_{+}\phi$ and $N=6$ for analyzed Yukawa model. The crucial point is, clearly, to perform the inverse matrix of the Dirac-Bergmann one. 

The matrix of the Poisson brackets of the constraints of the second class, involved in the equation (\ref{dirbracketdef}) as its inverse, is built of different structures. Firstly, the ${\cal F}(x^{+},{\bar w},{\bar z})$ has the standard algebraic form as the array. Secondly, each element of this matrix contains the distributions, their derivatives, the field operators or the functions of the continuous light-front coordinates. Thus, we encounter the case of the functional matrix, which requires the adjustment of the definition for its inverse. From this point forward, we take the following integral prescription 
\beq
\label{matrixintegralinversion} 
\int_{R^{3}}\!{d^{3}{\bar z}}\;
{\cal F}^{-1}\!\left(x^{+},{\bar x},{\bar z}\right)
{\cal F}\left(x^{+},{\bar z},{\bar y}\right)=
\delta^{(3)}\left({\bar x}-{\bar y}\right){\rm I}=
\int_{R^{3}}\!{d^{3}{\bar z}}\;
{\cal F}\left(x^{+},{\bar x},{\bar z}\right)
{\cal F}^{-1}\!\left(x^{+},{\bar z},{\bar y}\right). 
\eeq
The symbol ${\rm I}$ denotes the unit matrix of the relevant size, purely algebraic. Also the Dirac delta plays, as usual, role of the unit operator. It can be seen, the structure of the functional matrix ${\cal F}(x^{+},{\bar w},{\bar z})$ for the Yukawa model with higher order derivatives is rather complicated and therefore brings on some difficulties on the way of computing its inverse. The application of the Laplace method is not possible in this case at all. Nonetheless, the functional matrix ${\cal F}^{-1}(x^{+},{\bar x},{\bar y})$ may be easily and directly calculated for the case of free fermionic and scalar fields, when their interactions are switched off, namely $g=0=\lambda$, but the higher order derivative contribution still is present: $\alpha\neq0$. Therefore, we decompose discussed matrix 
${\cal F}(x^{+},{\bar w},{\bar z})$ in order to overcome the mathematical complications, as the sum of two other ones 
\beq
\label{fmatrixdecomp}
{\cal F}\left(x^{+},{\bar w},{\bar z}\right)=
{\cal F}_{0}\left(x^{+},{\bar w},{\bar z}\right)+
{\cal G}\left(x^{+},{\bar w},{\bar z}\right),
\eeq
where the first, ${\cal F}_{0}(x^{+},{\bar w},{\bar z})$, includes only the free fields, but embraces the higher order derivative contribution. The whole interactions of the model, including hard to master fermionic operators in the relevant vertex, are moved to the matrix 
${\cal G}(x^{+},{\bar w},{\bar z})$. We are in search of the algebraic formula, which allows to express the inverse sum of the matrices by the series of their inverses and themselves 
\beq
\label{matrixseries0}
\left(A+S\right)^{-1}=
A^{-1}+
\sum_{n=1}^{+\infty}
(-1)^{n}\left(A^{-1}S\right)^{n}\!A^{-1}, 
\eeq
whereas the $A^{-1}$ is not difficult to obtain, but the array $S^{-1}$ does not have to be known. To put it another way, the mathematical demands for existence of this equation contain invertibility of the whole matrix $A+S$ and also of the $A$. What's more, the second matrix, $S$, does not have to posses its inverse $S^{-1}$; it may be singular. This is our case, as we are about to see in the below considerations. Obviously, the functional extension of this pattern requires the summation with respect to the composite indexes, which are attached to each of our matrices and also entail relevant integrals with respect to the continuous space-time coordinates. 

Thus, if we are able to calculate the exact result for the inverse free matrix ${\cal F}^{-1}_{0}(x^{+},{\bar x},{\bar y})$, the general form of the ${\cal F}^{-1}(x^{+},{\bar x},{\bar y})$ is computed from the prescription (\ref{matrixseries0}), as the below chain of the matrices, where each product of them is assigned to the integration 
\beq
\label{matrixseries}
{\cal F}^{-1}\!\left(x^{+},{\bar x},{\bar y}\right)=
\left[
{\cal F}_{0}\left(x^{+},{\bar x},{\bar y}\right)+
{\cal G}\left(x^{+},{\bar x},{\bar y}\right)\right]^{-1}=
{\cal F}^{-1}_{0}\!\left(x^{+},{\bar x},{\bar y}\right)+
\eeq
\beqs
+\!\!\sum_{n=1}^{+\infty}\!
(-1)^{n}\!\!\!
\int_{R^{3}}\!\!\!\!\!\!{d^{3}{{\bar w}_{1}}}\!\!
\dots\!\!\!
\int_{R^{3}}\!\!\!\!\!\!{d^{3}{{\bar w}_{n}}}\!\!
\int_{R^{3}}\!\!\!\!\!\!{d^{3}{{\bar z}_{1}}}\!\!
\dots\!\!\!
\int_{R^{3}}\!\!\!\!\!\!{d^{3}{{\bar z}_{n}}}\!\!
\left\{\!
{\cal F}^{-1}_{0}\!\!\left(x^{+},{\bar x},{\bar w}_{1}\right)\!
{\cal G}\!\left(x^{+},{\bar w}_{1},{\bar z}_{1}\right)\!
{\cal F}^{-1}_{0}\!\!\left(x^{+},{\bar z}_{1},{\bar w}_{2}\right)\!
{\cal G}\!\left(x^{+},{\bar w}_{2},{\bar z}_{2}\right)\!\dots
\right.
\eeqs
\beqs
\left.
\dots{\cal F}^{-1}_{0}\!\!\left(x^{+},{\bar z}_{j-1},{\bar w}_{j}\right)\!
{\cal G}\!\left(x^{+},{\bar w}_{j},{\bar z}_{j}\right)\dots
{\cal F}^{-1}_{0}\!\!\left(x^{+},{\bar z}_{n-1},{\bar w}_{n}\right)\!
{\cal G}\!\left(x^{+},{\bar w}_{n},{\bar z}_{n}\right)\!
\right\}\!
{\cal F}^{-1}_{0}\!\left(x^{+},{\bar z}_{n},{\bar y}\right).
\eeqs
Accordingly, the decomposition (\ref{fmatrixdecomp}) leads in our case to the explicit expression for the free Dirac-Bergmann matrix  
\beq
\label{fmatrixfreeelements}
{\cal F}_{0}\left(x^{+},{\bar w},{\bar z}\right)=
\eeq
\beqs
\!=\!\!\left(\!
\begin{array}{cccccc}
0&\;\;0&\;\;0&\;\;0&\;\;0&\;\;
{\cal F}_{0}\left(x^{+},{\bar w},{\bar z}\right)_{16}\\
0&\;\;0&\;\;0&\;\;0&\;\;
{\cal F}_{0}\left(x^{+},{\bar w},{\bar z}\right)_{25}&\;\;
0\\
0&\;\;0&\;\;0&\;\;
{\cal F}_{0}\left(x^{+},{\bar w},{\bar z}\right)_{34}&\;\;
0&\;\;0\\
0&\;\;0&\;\;
{\cal F}_{0}\left(x^{+},{\bar w},{\bar z}\right)_{43}&\;\;
{\cal F}_{0}\left(x^{+},{\bar w},{\bar z}\right)_{44}&\;\; 
0&\;\;0\\
0&\;\;
{\cal F}_{0}\left(x^{+},{\bar w},{\bar z}\right)_{52}&\;\; 
0&\;\;0&\;\;0&\;\; 
{\cal F}_{0}\left(x^{+},{\bar w},{\bar z}\right)_{56}\\ 
{\cal F}_{0}\left(x^{+},{\bar w},{\bar z}\right)_{61}&\;\;
0&\;\;0&\;\;0&\;\;
{\cal F}_{0}\left(x^{+},{\bar w},{\bar z}\right)_{65}&\;\;
0\\
\end{array}
\!\!\right)\!\!.
\eeqs  
The elements of this matrix are specified below: 
\beq
\label{freemx1}
{\cal F}_{0}\left(x^{+},{\bar w},{\bar z}\right)_{16}=
2i\partial^{w}_{-}
\delta^{(3)}\left({\bar w}-{\bar z}\right)\Lambda_{-}, \;\;\;\;\;\;\;\;\;    
{\cal F}_{0}\left(x^{+},{\bar w},{\bar z}\right)_{25}=
2i\partial^{w}_{-}
\delta^{(3)}\left({\bar w}-{\bar z}\right)\Lambda_{-}, 
\eeq
\beq
\label{freemx2}
{\cal F}_{0}\left(x^{+},{\bar w},{\bar z}\right)_{34}=
-2{\alpha}\partial^{2w}_{-}
\delta^{(3)}\left({\bar w}-{\bar z}\right){\rm P}, \;\;\;\;\;\;\;\;\;   
{\cal F}_{0}\left(x^{+},{\bar w},{\bar z}\right)_{43}=
2{\alpha}\partial^{2w}_{-}
\delta^{(3)}\left({\bar w}-{\bar z}\right){\rm P}, 
\eeq
\beq
\label{freemx3}
{\cal F}_{0}\left(x^{+},{\bar w},{\bar z}\right)_{44}=
-2\partial^{w}_{-}
\delta^{(3)}\left({\bar w}-{\bar z}\right){\rm P}-
2{\alpha}\partial^{w}_{-}
\Delta^{w}_{\perp}\delta^{(3)}\left({\bar w}-{\bar z}\right){\rm P}, 
\eeq
\beq
\label{freemx4}
{\cal F}_{0}\left(x^{+},{\bar w},{\bar z}\right)_{52}=
-2i\partial^{w}_{-}
\delta^{(3)}\left({\bar w}-{\bar z}\right)\Lambda_{-}, \;\;\;\;\;\;\;\;\;  
{\cal F}_{0}\left(x^{+},{\bar w},{\bar z}\right)_{56}=
{\xi}_{0}\left(x^{+},{\bar w},{\bar z}\right)
\delta^{(3)}\left({\bar w}-{\bar z}\right),
\eeq
\beq
\label{freemx5}
{\cal F}_{0}\left(x^{+},{\bar w},{\bar z}\right)_{61}=
-2i\partial^{w}_{-}
\delta^{(3)}\left({\bar w}-{\bar z}\right)\Lambda_{-}, \;\;\;\;\;\;\;\;\; 
{\cal F}_{0}\left(x^{+},{\bar w},{\bar z}\right)_{65}=
-{\xi}_{0}\left(x^{+},{\bar w},{\bar z}\right)
\delta^{(3)}\left({\bar w}-{\bar z}\right).  
\eeq
The differential operator $\xi_{0}(x^{+},{\bar w},{\bar z})$ states the free part of the $\xi(x^{+},{\bar w},{\bar z})$, defined by the pattern (\ref{amatrixfunction}), in which the Yukawa coupling constant $g$ is taken as zero 
\beq
\label{amatrixfunctionzero}
{\xi}_{0}\left(x^{+},{\bar w},{\bar z}\right)=
i\sqrt{2}\Delta^{w}_{\perp}\Lambda_{-}-
i\sqrt{2}M^{2}\Lambda_{-}. 
\eeq  
The comprehensive information about the structure of the whole interactions in this model is involved in the matrix 
${\cal G}(x^{+},{\bar w},{\bar z})$. This contains the fermionic operators and yields 
\beq
\label{gmatrixelements}
{\cal G}\left(x^{+},{\bar w},{\bar z}\right)=
\left(\;
\begin{array}{cccccc}
0&\;\;0&\;\;0&\;\;0&\;\;0&\;\;0\\
0&\;\;0&\;\;0&\;\;0&\;\;0&\;\;0\\
0&\;\;0&\;\;0&\;\;0&\;\;0&\;\;0\\
0&\;\;0&\;\;0&\;\;0&\;\;
{\cal G}_{45}\left(x^{+},{\bar w},{\bar z}\right)&\;\;
{\cal G}_{46}\left(x^{+},{\bar w},{\bar z}\right)\\
0&\;\;0&\;\;0&\;\;
{\cal G}_{54}\left(x^{+},{\bar w},{\bar z}\right)&\;\;
0&\;\; 
{\cal G}_{56}\left(x^{+},{\bar w},{\bar z}\right)\\
0&\;\;0&\;\;0&\;\;
{\cal G}_{64}\left(x^{+},{\bar w},{\bar z}\right)&\;\;
{\cal G}_{65}\left(x^{+},{\bar w},{\bar z}\right)&\;\;
0\\
\end{array}
\right).  
\eeq
The elements of the above ${\cal G}(x^{+},{\bar w},{\bar z})$ satisfy:
\beq
\label{gmatrix1}
{\cal G}_{45}\left(x^{+},{\bar w},{\bar z}\right)=
-g\Psi^{\dagger}_{+}\left(x^{+},{\bar z}\right)\gamma^{-}
\delta^{(3)}\left({\bar w}-{\bar z}\right), \;\;\;\;\;\; 
{\cal G}_{46}\left(x^{+},{\bar w},{\bar z}\right)=
g\gamma^{+}\Psi_{+}\left(x^{+},{\bar z}\right)
\delta^{(3)}\left({\bar w}-{\bar z}\right),
\eeq
\beq
\label{gmatrix2}
{\cal G}_{54}\left(x^{+},{\bar w},{\bar z}\right)=
g\Psi^{\dagger}_{+}\left(x^{+},{\bar w}\right)\gamma^{-}
\delta^{(3)}\left({\bar w}-{\bar z}\right), \;\;\;\;\;\; 
{\cal G}_{56}\left(x^{+},{\bar w},{\bar z}\right)=
{\xi}_{g}\left(x^{+},{\bar w},{\bar z}\right)
\delta^{(3)}\left({\bar w}-{\bar z}\right),
\eeq
\beq
\label{gmatrix3}
{\cal G}_{64}\left(x^{+},{\bar w},{\bar z}\right)=
-g\gamma^{+}\Psi_{+}\left(x^{+},{\bar w}\right)
\delta^{(3)}\left({\bar w}-{\bar z}\right), \;\;\;\;\;\; 
{\cal G}_{65}\left(x^{+},{\bar w},{\bar z}\right)=
-{\xi}_{g}\left(x^{+},{\bar w},{\bar z}\right)
\delta^{(3)}\left({\bar w}-{\bar z}\right).
\eeq
The new object $\xi_{g}(x^{+},{\bar w},{\bar z})=\xi(x^{+},{\bar w},{\bar z})-\xi_{0}(x^{+},{\bar w},{\bar z})$ is introduced as the operator, which includes the pure term of the interacting scalar field 
\beq
\label{amatrixfunctioninteract}
{\xi}_{g}\left(x^{+},{\bar w},{\bar z}\right)=
-i\sqrt{2}gM
\left[
{\phi}\left(x^{+},{\bar w}\right)+{\phi}\left(x^{+},{\bar z}\right)
\right]
\Lambda_{-}-
i\sqrt{2}g^{2}
{\phi}\left(x^{+},{\bar w}\right){\phi}\left(x^{+},{\bar z}\right)\Lambda_{-}. 
\eeq 
Now, if we take the expansion (\ref{matrixseries}) as the point of departure, we have two slightly different approaches to the further computations. The equation (\ref{matrixseries}) represents finite case for when, the matrix series on the right side is truncated with regard to the zeroth value of the product of the relevant matrices. If we are able to perform the infinite summation on the right side of the above formula, we also obtain the finite results. This is the first approach, for which is more suitable to use the pure matrix expansion, despite the structure of the right-sided terms, proportional to the different powers of the coupling constant $g$. This represents the exact solution for our problem. For the infinite expansion, where we cannot dispose the general solution of the equation (\ref{matrixseries}), we may regard, if it is physically reasonable, the matrix $S$ as comprising a small parameter - the dimensionless coupling constant $g\ll{1}$, for instance. It sometimes allows to run effective calculations, physically treated as the perturbations. This is the second approach. Obviously, both these methods are fully equivalent. 

We can see immediately, the first approach is just described by the prescription (\ref{matrixseries}). This may only be rewritten for the convenience in more compact form 
\beq
\label{finverseperturbative0} 
{\cal F}^{-1}\left(x^{+},{\bar x},{\bar y}\right)=
{\cal F}^{-1}_{0}\left(x^{+},{\bar x},{\bar y}\right)+
\sum_{n=1}^{+\infty}
{\cal A}_{n}\left(x^{+},{\bar x},{\bar y}\right).    
\eeq 
Each contribution ${\cal A}_{n}(x^{+},{\bar x},{\bar y})$ of the $n$th order to the above series represents the relevant product of the free matrix ${\cal F}^{-1}_{0}$ and already introduced array with interactions ${\cal G}$. This is 
\beq
\label{finverseperturbativen0}
{\cal A}_{n}\left(x^{+},{\bar x},{\bar y}\right)=
\eeq
\beqs
=(-1)^{n}
\left\{
{\prod_{j=1}^{n}}
\left[
\int_{R^{3}}\!\!{d^{3}{\bar w}_{j}}\!\!
\int_{R^{3}}\!\!{d^{3}{\bar z}_{j}}\;\!
{\cal F}^{-1}_{0}\!\left(x^{+},{\bar z}_{j-1},{\bar w}_{j}\right)
{\cal G}\left(x^{+},{\bar w}_{j},{\bar z}_{j}\right)
\right]_{{{\bar{z}}_{0}={\bar x}}}
\right\}
{\cal F}^{-1}_{0}\!\left(x^{+},{\bar z}_{n},{\bar y}\right).
\eeqs
Noticeably, both powers, the first and the second, of the Yukawa coupling constant $g$ are included in the matrix ${\cal G}$. Therefore, the constant $g$ is not a parameter of the expansion. 

The second approach, devoted to obtaining the inverse matrix ${\cal F}^{-1}(x^{+},{\bar x},{\bar y})$, relies on the decomposition of the matrix ${\cal G}(x^{+},{\bar w},{\bar z})$. This array embraces the elements, which are determined by the terms proportional to the first and to the second powers of the coupling constant $g$. On account of that, we can decompose the ${\cal G}(x^{+},{\bar w},{\bar z})$ into the sum of two matrix contributions, proportional to aforementioned $g$ and $g^{2}$. This finally allows us to write 
\beq
\label{matrixg12}
{\cal G}\left(x^{+},{\bar w},{\bar z}\right)=
g\;\!{\tilde{\cal G}}_{1}\left(x^{+},{\bar w},{\bar z}\right)+
g^{2}\;\!{\tilde{\cal G}}_{2}\left(x^{+},{\bar w},{\bar z}\right).
\eeq
Thus, we have established two matrices, the ${\tilde{\cal G}}_{1}(x^{+},{\bar w},{\bar z})$ and the 
${\tilde{\cal G}}_{2}(x^{+},{\bar w},{\bar z})$. The first one satisfies 
\beq
\label{gtildematrixelement1}
{\tilde{\cal G}}_{1}\left(x^{+},{\bar w},{\bar z}\right)=
\left(\;
\begin{array}{cccccc}
0&\;\;0&\;\;0&\;\;0&\;\;0&\;\;0\\
0&\;\;0&\;\;0&\;\;0&\;\;0&\;\;0\\
0&\;\;0&\;\;0&\;\;0&\;\;0&\;\;0\\
0&\;\;0&\;\;0&\;\;0&\;\; 
{\tilde{\cal G}}_{1}\left(x^{+},{\bar w},{\bar z}\right)_{45}&\;\;
{\tilde{\cal G}}_{1}\left(x^{+},{\bar w},{\bar z}\right)_{46}\\
0&\;\;0&\;\;0&\;\; 
{\tilde{\cal G}}_{1}\left(x^{+},{\bar w},{\bar z}\right)_{54}&\;\;
0&\;\;
{\tilde{\cal G}}_{1}\left(x^{+},{\bar w},{\bar z}\right)_{56}\\ 
0&\;\;0&\;\;0&\;\;
{\tilde{\cal G}}_{1}\left(x^{+},{\bar w},{\bar z}\right)_{64}&\;\;
{\tilde{\cal G}}_{1}\left(x^{+},{\bar w},{\bar z}\right)_{65}&\;\;
0\\
\end{array}
\right).
\eeq
The elements of the above array are as follows: 
\beq
\label{g1matrix1}
{\tilde{\cal G}}_{1}\left(x^{+},{\bar w},{\bar z}\right)_{45}=
-\Psi^{\dagger}_{+}\left(x^{+},{\bar z}\right)\gamma^{-}
\delta^{(3)}\left({\bar w}-{\bar z}\right), \;\;\;\;\;\; 
{\tilde{\cal G}}_{1}\left(x^{+},{\bar w},{\bar z}\right)_{46}=   
\gamma^{+}\Psi_{+}\left(x^{+},{\bar z}\right)
\delta^{(3)}\left({\bar w}-{\bar z}\right),
\eeq
\beq
\label{g1matrix2}
{\tilde{\cal G}}_{1}\left(x^{+},{\bar w},{\bar z}\right)_{54}=
\Psi^{\dagger}_{+}\left(x^{+},{\bar w}\right)\gamma^{-}
\delta^{(3)}\left({\bar w}-{\bar z}\right), \;\;\;\;\;\; 
{\tilde{\cal G}}_{1}\left(x^{+},{\bar w},{\bar z}\right)_{56}=
{\tilde\xi}_{1}\left(x^{+},{\bar w},{\bar z}\right)
\delta^{(3)}\left({\bar w}-{\bar z}\right), 
\eeq
\beq
\label{g1matrix3}
{\tilde{\cal G}}_{1}\left(x^{+},{\bar w},{\bar z}\right)_{64}= 
-\gamma^{+}\Psi_{+}\left(x^{+},{\bar w}\right)
\delta^{(3)}\left({\bar w}-{\bar z}\right), \;\;\;\;\;\; 
{\tilde{\cal G}}_{1}\left(x^{+},{\bar w},{\bar z}\right)_{65}=
-{\tilde\xi}_{1}\left(x^{+},{\bar w},{\bar z}\right)
\delta^{(3)}\left({\bar w}-{\bar z}\right). 
\eeq
The second matrix ${\tilde{\cal G}}_{2}(x^{+},{\bar w},{\bar z})$ is already having the simple form
\beq
\label{gtildematrixelement2}
{\tilde{\cal G}}_{2}\left(x^{+},{\bar w},{\bar z}\right)=
\left(\;
\begin{array}{cccccc}
0&\;\;0&\;\;0&\;\;0&\;\;0&\;\;0\\
0&\;\;0&\;\;0&\;\;0&\;\;0&\;\;0\\
0&\;\;0&\;\;0&\;\;0&\;\;0&\;\;0\\
0&\;\;0&\;\;0&\;\;0&\;\;0&\;\;0\\
0&\;\;0&\;\;0&\;\;0&\;\;0&\;\; 
{\tilde{\cal G}}_{2}\left(x^{+},{\bar w},{\bar z}\right)_{56}\\ 
0&\;\;0&\;\;0&\;\;0&\;\; 
{\tilde{\cal G}}_{2}\left(x^{+},{\bar w},{\bar z}\right)_{65}&\;\;
0\\
\end{array}
\right),
\eeq   
wherein the elements obey:
\beq
\label{g2matrixall}
{\tilde{\cal G}}_{2}\left(x^{+},{\bar w},{\bar z}\right)_{56}=
{\tilde\xi}_{2}\left(x^{+},{\bar w},{\bar z}\right)
\delta^{(3)}\left({\bar w}-{\bar z}\right), \;\;\;\;\;\; 
{\tilde{\cal G}}_{2}\left(x^{+},{\bar w},{\bar z}\right)_{65}=
-{\tilde\xi}_{2}\left(x^{+},{\bar w},{\bar z}\right)
\delta^{(3)}\left({\bar w}-{\bar z}\right). 
\eeq
Both the operators ${\tilde\xi}_{1}(x^{+},{\bar w},{\bar z})$ and ${\tilde\xi}_{2}(x^{+},{\bar w},{\bar z})$ constitute already defined $\xi_{g}(x^{+},{\bar w},{\bar z})$, due to the power decomposition (\ref{amatrixfunctioninteract}) of this part of the Dirac-Bergmann matrix, which includes the interactions: 
\beq
\label{xidecomp}
{\xi}_{g}\!\left(x^{+},{\bar w},{\bar z}\right)=
g{\tilde\xi}_{1}\!\left(x^{+},{\bar w},{\bar z}\right)+
g^{2}{\tilde\xi}_{2}\!\left(x^{+},{\bar w},{\bar z}\right),
\eeq
\beq
\label{ksi12}
{\tilde\xi}_{1}\!\left(x^{+},{\bar w},{\bar z}\right)\!=\!
-i\sqrt{2}M\!
\left[
{\phi}\left(x^{+},{\bar w}\right)\!+\!
{\phi}\left(x^{+},{\bar z}\right)
\right]\!
\Lambda_{-}, \;\;\;\;\;\; 
{\tilde\xi}_{2}\!\left(x^{+},{\bar w},{\bar z}\right)\!=\!
-i\sqrt{2}
{\phi}\left(x^{+},{\bar w}\right)
{\phi}\left(x^{+},{\bar z}\right)\!
\Lambda_{-}. 
\eeq 
The expansion (\ref{matrixseries}) of the inverse matrix ${\cal F}^{-1}(x^{+},{\bar x},{\bar y})$ should now be re-expressed by way of the effective pattern \cite{b20}. This is possible by the insertion of the decomposition of the ${\cal G}(x^{+},{\bar w},{\bar z})$ array into two terms (\ref{matrixg12}) with relevant powers of the coupling constant $g$. As a consequence, there the series occurs with sequent powers of the said $g$. Finally 
\beq
\label{finverseperturbative} 
{\cal F}^{-1}\left(x^{+},{\bar x},{\bar y}\right)=
{\cal F}^{-1}_{0}\left(x^{+},{\bar x},{\bar y}\right)+
\sum_{n=1}^{+\infty}
{\cal B}_{n}\left(x^{+},{\bar x},{\bar y}\right),   
\eeq
where the successive orders of obtained perturbations are:  
\beq
\label{finverseperturbativen}
{\cal B}_{n}\left(x^{+},{\bar x},{\bar y}\right)=
\eeq
\beqs
=g^{n}\!\!\!
\sum_{{{\bar{\rm{V}}}^{1,\dots,n}}_{2}}
\sum_{l=1}^{n}
(-1)^{l}\!\!\;\!
\left\{
{\prod_{j=1}^{l}}{{ }^{{ }^{\prime}}}\!
\left[
\int_{R^{3}}\!\!{d^{3}{\bar w}_{j}}\!\!
\int_{R^{3}}\!\!{d^{3}{\bar z}_{j}}\;\!
{\cal F}^{-1}_{0}\!\left(x^{+},{\bar z}_{j-1},{\bar w}_{j}\right)
{\tilde{\cal G}}_{{\bar{\rm{v}}}_{2}^{j}}\!\left(x^{+},{\bar w}_{j},{\bar z}_{j}\right)
\right]_{{{\bar z}_{0}={\bar{x}}}}\!
\right\}\!
{\cal F}^{-1}_{0}\!\left(x^{+},{\bar z}_{l},{\bar y}\right).
\eeqs
The first sum runs all the elements of the set ${\bar{\rm{V}}}^{1,\dots,n}_{2}={\bar{\rm{V}}}^{1}_{2}\cup\dots\cup{\bar{\rm{V}}}^{n}_{2}$, which is the union of the sets ${\bar{\rm{V}}}^{r}_{2}$, provided that each of them separately includes the maps being the $r$-tuples of the two-element set ${\bar{\rm{v}}}^{r}_{2}\in{\bar{\rm{V}}}^{r}_{2}$ and $r=1,\dots,n$. The prime next to the symbol of the product shows, that the sum of the indexes in the chain of the matrices ${\tilde{\cal G}}_{1,2}$ is ${\bar{\rm{v}}}^{1}_{2}+\dots+{\bar{\rm{v}}}^{l}_{2}=n$, wherein the ${\bar{\rm{v}}}^{j}_{2}$ denotes not only the $j$-tuple of the two-element set $\{1,2\}$, but simultaneously means the value of this map: ${\bar{\rm{v}}}^{j}_{2}=1,2$, whilst $j=1,\dots,l$. This condition provides in fact, that the sum of the indexes of the aforesaid matrices ${\tilde{\cal G}}_{1,2}$ gives the condition for their chain length or more precisely for their number $l$ within each contribution to 
the (\ref{finverseperturbativen}), determining the $n$th fixed order of the series ${\cal B}_{n}(x^{+},{\bar x},{\bar y})$. 

\section{Free Inverse Dirac-Bergmann Matrix}
\label{ifmf}

We should commence searches of the free inverse matrix of the Dirac-Bergmann one from below remarks of the mathematical nature. We observe, that the structure of the matrix (\ref{fmatrixfreeelements}) is fully anti-down-triangle. Therefore, we postulate, that its inverse has the form of the anti-up-triangle array, but the small block of the size $2\times2$ at the very center is modified by the non-trivial element 
${\cal F}^{-1}_{044}(x^{+},{\bar x}-{\bar y})$. This one is necessary for the reduction of our model to the case of $\alpha=0$. Finally, the inverse matrix is the reflection with respect to the anti-diagonal of the initial array ${\cal F}_{0}(x^{+},{\bar w},{\bar z})$, but not pure. Until this moment we can say, in our case, about the well defined procedure of obtaining the inverse of the functional matrix. In general, the computation of the functional inverse matrix may lead if its algebraic frame is not properly introduced, to many ambiguities. Some modifications of this algebraic frame are acceptable if there is another reason for it, like in this instance, where we add the non-trivial element ${\cal F}^{-1}_{044}(x^{+},{\bar x}-{\bar y})$. Therefore, we postulate the inverse of the free Dirac-Bergmann matrix (\ref{fmatrixfreeelements}) as 
\beq
\label{invfmatrixelements} 
{\cal F}^{-1}_{0}\left(x^{+},{\bar x},{\bar y}\right)=
{\cal F}^{-1}_{0}\left(x^{+},{\bar x}-{\bar y}\right)=
\eeq
\beqs
=\!\left(
\begin{array}{cccccc}
0&\; 
{\cal F}^{\!\!\mbox{ }_{-1}}_{\!\!012}\!\!\left(x^{+}\!,{\bar x}-{\bar y}\right)&\;
0&\;0&\;0&\;
{\cal F}^{\!\!\mbox{ }_{-1}}_{\!\!016}\!\!\left(x^{+}\!,{\bar x}-{\bar y}\right)\\
{\cal F}^{\!\!\mbox{ }_{-1}}_{\!\!021}\!\!\left(x^{+}\!,{\bar x}-{\bar y}\right)&\;
0&\;0&\;0&\;
{\cal F}^{\!\!\mbox{ }_{-1}}_{\!\!025}\!\!\left(x^{+}\!,{\bar x}-{\bar y}\right)&\;
0\\
0&\;0&\;
{\cal F}^{\!\!\mbox{ }_{-1}}_{\!\!033}\!\!\left(x^{+}\!,{\bar x}-{\bar y}\right)&\;
{\cal F}^{\!\!\mbox{ }_{-1}}_{\!\!034}\!\!\left(x^{+}\!,{\bar x}-{\bar y}\right)&\; 
0&\;0\\
0&\;0&\;
{\cal F}^{\!\!\mbox{ }_{-1}}_{\!\!043}\!\!\left(x^{+}\!,{\bar x}-{\bar y}\right)&\;
{\cal F}^{\!\!\mbox{ }_{-1}}_{\!\!044}\!\!\left(x^{+}\!,{\bar x}-{\bar y}\right)&\; 
0&\;0\\
0&\;
{\cal F}^{\!\!\mbox{ }_{-1}}_{\!\!052}\!\!\left(x^{+}\!,{\bar x}-{\bar y}\right)&\;
0&\;0&\;0&\;0\\ 
{\cal F}^{\!\!\mbox{ }_{-1}}_{\!\!061}\!\!\left(x^{+}\!,{\bar x}-{\bar y}\right)&\;
0&\;0&\;0&\;0&\;0\\
\end{array}
\right)\!.   
\eeqs
We introduced the elements of this inverse array as the following functions to determine: 
\beq
\label{invfmatrixelements1} 
{\cal F}^{-1}_{012}\left(x^{+},{\bar x}-{\bar y}\right)=
c\left(x^{+},{\bar x}-{\bar y}\right)\Lambda_{-}, \;\;\;\;\;\;\;\;\;  
{\cal F}^{-1}_{016}\left(x^{+},{\bar x}-{\bar y}\right)=
-a\left(x^{+},{\bar x}-{\bar y}\right)\Lambda_{-}, 
\eeq
\beq
\label{invfmatrixelements2} 
{\cal F}^{-1}_{021}\left(x^{+},{\bar x}-{\bar y}\right)=
-c\left(x^{+},{\bar x}-{\bar y}\right)\Lambda_{-}, \;\;\;\;\;\;\;\;\;   
{\cal F}^{-1}_{025}\left(x^{+},{\bar x}-{\bar y}\right)=
-a\left(x^{+},{\bar x}-{\bar y}\right)\Lambda_{-},
\eeq
\beq
\label{invfmatrixelements3} 
{\cal F}^{-1}_{033}\left(x^{+},{\bar x}-{\bar y}\right)=
b_{1}\left(x^{+},{\bar x}-{\bar y}\right){\rm P}, \;\;\;\;\;\;\;\;\;  
{\cal F}^{-1}_{034}\left(x^{+},{\bar x}-{\bar y}\right)=
d_{1}\left(x^{+},{\bar x}-{\bar y}\right){\rm P}, 
\eeq
\beq
\label{invfmatrixelements4} 
{\cal F}^{-1}_{043}\left(x^{+},{\bar x}-{\bar y}\right)=
d_{2}\left(x^{+},{\bar x}-{\bar y}\right){\rm P}, \;\;\;\;\;\;\;\;\; 
{\cal F}^{-1}_{044}\left(x^{+},{\bar x}-{\bar y}\right)=
b_{2}\left(x^{+},{\bar x}-{\bar y}\right){\rm P}, 
\eeq
\beq
\label{invfmatrixelements5} 
{\cal F}^{-1}_{052}\left(x^{+},{\bar x}-{\bar y}\right)=
a\left(x^{+},{\bar x}-{\bar y}\right)\Lambda_{-}, \;\;\;\;\;\;\;\;\;  
{\cal F}^{-1}_{061}\left(x^{+},{\bar x}-{\bar y}\right)=
a\left(x^{+},{\bar x}-{\bar y}\right)\Lambda_{-}. 
\eeq
The matrix ${\cal F}^{-1}_{0}(x^{+},{\bar x}-{\bar y})$ does not embrace the interactions and therefore is translationally symmetric, what explicitly manifests in the second argument of its elements, being the subtractions of the relevant light-front coordinates. 

Both, the left- and the right-sided conditions (\ref{matrixintegralinversion}) of the unambiguous existence of the inverse matrix, applied to our case (\ref{invfmatrixelements}), lead to the system of the differential equations for the above functions. The beginning and the primary form of this system, taken directly from mentioned conditions, yields: 
\beq
\label{acdiffeq}
2i\partial_{-}a\left(x^{+},{\bar x}\right)=
\delta^{(3)}\left({\bar x}\right), \;\;\;\;\;\;\;\;\;   
2\partial_{-}c\left(x^{+},{\bar x}\right)+
\sqrt{2} 
\left(\Delta_{\perp}-M^{2}\right)
a\left(x^{+},{\bar x}\right)
=0,
\eeq
\beq
\label{b1b1diffeq}
\alpha\partial^{2}_{-}b_{1}\left(x^{+},{\bar x}\right)+ 
\partial_{-}\left(\alpha\Delta_{\perp}+1\right)
d_{1}\left(x^{+},{\bar x}\right)
=0, \;\;\;\;\;\; 
\alpha\partial^{2}_{-}b_{1}\left(x^{+},{\bar x}\right)- 
\partial_{-}\left(\alpha\Delta_{\perp}+1\right)
d_{2}\left(x^{+},{\bar x}\right)
=0, 
\eeq
\beq
\label{b2diffeq}
\alpha\partial^{2}_{-}
b_{2}\left(x^{+},{\bar x}\right)
=0,
\eeq
\beq
\label{d1d1diffeq}
2\alpha\partial^{2}_{-}
d_{1}\left(x^{+},{\bar x}\right)=
\delta^{(3)}\left({\bar x}\right), \;\;\;\;\;\;\;\;\; 
2\alpha\partial^{2}_{-}d_{1}\left(x^{+},{\bar x}\right)-
2\partial_{-}\left(\alpha\Delta_{\perp}+1\right)
b_{2}\left(x^{+},{\bar x}\right)=
\delta^{(3)}\left({\bar x}\right), 
\eeq
\beq
\label{d2d2diffeq}
-2\alpha\partial^{2}_{-}
d_{2}\left(x^{+},{\bar x}\right)=
\delta^{(3)}\left({\bar x}\right), \;\;\;\;\;\;\;\;\;  
-2\alpha\partial^{2}_{-}d_{2}\left(x^{+},{\bar x}\right)-
2\partial_{-}\left(\alpha\Delta_{\perp}+1\right)
b_{2}\left(x^{+},{\bar x}\right)=
\delta^{(3)}\left({\bar x}\right).  
\eeq
This can be brought, after the elementary transformations, to: 
\beq
\label{acdiffeqfin}
2i\partial_{-}a\left(x^{+},{\bar x}\right)=
\delta^{(3)}\left({\bar x}\right), \;\;\;\;\;\;\;\;\;  
2\partial_{-}c\left(x^{+},{\bar x}\right)+
\sqrt{2}\left(\Delta_{\perp}-M^{2}\right)
a\left(x^{+},{\bar x}\right)
=0,
\eeq
\beq
\label{d1d2diffeqfin}
2\alpha\partial^{2}_{-}
d_{1}\left(x^{+},{\bar x}\right)=
\delta^{(3)}\left({\bar x}\right), \;\;\;\;\;\;\;\;\;   
-2\alpha\partial_{-}d_{2}\left(x^{+},{\bar x}\right)=
\delta^{(3)}\left({\bar x}\right), 
\eeq
\beq
\label{b1b1diffeqfin}
\alpha\partial^{2}_{-}b_{1}\left(x^{+},{\bar x}\right)+ 
\partial_{-}\left(\alpha\Delta_{\perp}+1\right)
d_{1}\left(x^{+},{\bar x}\right)
=0, \;\;\;\;\;\; 
\alpha\partial^{2}_{-}b_{1}\left(x^{+},{\bar x}\right)- 
\partial_{-}\left(\alpha\Delta_{\perp}+1\right)
d_{2}\left(x^{+},{\bar x}\right)
=0,  
\eeq
\beq
\label{b2b2diffeqfin}
\alpha\partial^{2}_{-}
b_{2}\left(x^{+},{\bar x}\right)
=0, \;\;\;\;\;\;\;\;\;\;\;\;  
\partial_{-}\left(\alpha\Delta_{\perp}+1\right)
b_{2}\left(x^{+},{\bar x}\right)
=0. 
\eeq
We have three separate subsystems of these equations: the (\ref{acdiffeqfin}), the (\ref{b1b1diffeqfin}) in relation to (\ref{d1d2diffeqfin}) and the (\ref{b2b2diffeqfin}). We easily observe, that the first (\ref{acdiffeqfin}) and the third (\ref{b2b2diffeqfin}) ones are consistent and not overdetermined. We have to conclude, by comparing the (\ref{b1b1diffeqfin}) with both the equations (\ref{d1d2diffeqfin}), that this subsystem of the equations for the function $b_{1}(x^{+},{\bar x})$ is consistent and not overdetermined, under the condition 
\beq
\label{diffsystcond}
\partial_{-}
d_{2}\left(x^{+},{\bar x}\right)=
-\partial_{-}
d_{1}\left(x^{+},{\bar x}\right). 
\eeq
We should take this into account by solving, say, the second one of the equations (\ref{d1d2diffeqfin}). We conclude, by incorporating the addressed condition, that the system of the equations (\ref{acdiffeq}) - (\ref{d2d2diffeq}) and these being equivalent or deduced from this, are also consistent. From this moment on the subsystem (\ref{b1b1diffeqfin}) includes only one independent equation. 

The general solution for the system of the equations (\ref{acdiffeqfin}) - (\ref{b2b2diffeqfin}) with the condition (\ref{diffsystcond}) in the domain of the distribution should not cause serious difficulties: 
\beq
\label{asolut}
a\left({\bar x}\right)=
{1\over{4i}}\;
{\rm sgn}\left(x^{-}\right)
\delta^{(2)}\left({\bf x}_{\perp}\right)+
a_{0}\left({\bf x}_{\perp}\right), 
\eeq
\beq
\label{csolut}
c\left({\bar x}\right)=
-{1\over{4\sqrt{2}i}}
\mid\!{x^{-}}\!\mid
\left(\Delta_{\perp}-M^{2}\right)
\delta^{(2)}\left({\bf x}_{\perp}\right)-
{1\over{\sqrt{2}}}\;x^{-}
\left(\Delta_{\perp}-M^{2}\right)
a_{0}\left({\bf x}_{\perp}\right)+
c_{0}\left({\bf x}_{\perp}\right), 
\eeq
\beq
\label{d1solut}
d_{1}\left({\bar x}\right)=
{1\over{4\alpha}}
\mid\!{x^{-}}\!\mid
\delta^{(2)}\left({\bf x}_{\perp}\right)+
x^{-}d_{10}\left({\bf x}_{\perp}\right)+
d_{100}\left({\bf x}_{\perp}\right),
\eeq
\beq
\label{d2solut}
d_{2}\left({\bar x}\right)=
-{1\over{4\alpha}}
\mid\!{x^{-}}\!\mid
\delta^{(2)}\left({\bf x}_{\perp}\right)-
x^{-}d_{10}\left({\bf x}_{\perp}\right)+
d_{200}\left({\bf x}_{\perp}\right),
\eeq
\beq
\label{b1solut}
b_{1}\left({\bar x}\right)=
\eeq
\beqs
=-{1\over{8\alpha}}
{\rm sgn}\left(x^{-}\right)
\left(x^{-}\right)^{2}
\left(\Delta_{\perp}+{1\over{\alpha}}\right)
\delta^{(2)}\left({\bf x}_{\perp}\right)-
{1\over 2}
\left(x^{-}\right)^{2}
\left(\Delta_{\perp}+{1\over{\alpha}}\right)
d_{10}\left({\bf x}_{\perp}\right)+
x^{-}b_{10}\left({\bf x}_{\perp}\right)+
b_{100}\left({\bf x}_{\perp}\right)\!, 
\eeqs
\beq
\label{b2solut}
b_{2}\left({\bar x}\right)=
x^{-}B\left({\bf x}_{\perp}\right)+
b_{20}\left({\bf x}_{\perp}\right).  
\eeq
We have the following differential equation for the $B({\bf x}_{\perp})$ function 
\beq
\label{Bsolut}
\left(\Delta_{\perp}+{1\over{\alpha}}\right)
B\left({\bf x}_{\perp}\right)
=0,
\eeq
inferred from the second pattern of the subsystem (\ref{b2b2diffeqfin}). The most general solution of this eigenproblem is well known in many coordinate systems. At this point we confine ourselves to quote this obvious fact, until then without broader discussion, concluding only, that necessary is to exclude the case $B({\bf x}_{\perp})={constant}\neq0$, as not satisfying the (\ref{Bsolut}). The solution $B({\bf x}_{\perp})=0$ is acceptable, but together with the requirement $b_{20}({\bf x}_{\perp})\neq0$, due to the fact, that we introduced the non-trivial function $b_{2}({\bar x})$ for reduction of our problem to the case of $\alpha=0$. The latter describes the pure Yukawa model, discussed from presented point of view in \cite{b20}. 

Of course, the general solutions (\ref{asolut}) - (\ref{b2solut}) are not uniquely determined. The functions: $a_{0}({\bf x}_{\perp})$,  $b_{10}({\bf x}_{\perp})$, $b_{100}({\bf x}_{\perp})$, $b_{20}({\bf x}_{\perp})\neq0$, $c_{0}({\bf x}_{\perp})$, $d_{10}({\bf x}_{\perp})$,  $d_{100}({\bf x}_{\perp})$ and $d_{200}({\bf x}_{\perp})$ are for now entirely arbitrary. Just like the $B({\bf x}_{\perp})$, which however represents the solution of the eigenproblem (\ref{Bsolut}), but requires the boundary conditions for its final obtaining. Aforementioned boundary conditions, imposed on the studied solutions (\ref{asolut}) - (\ref{b2solut}), should transform them into unequivocally determined. They also allow the computation of the unambiguous light-front Dirac brackets. From the other side, there are many lacks of the clarity, referred to the physical correctness of these boundary conditions and relevant to the violation of the Lorentz or the internal symmetries of the analyzed model \cite{b21}. Now, to avoid all discussed ambiguities, we make the precondition, that physically acceptable is the simplest case - the special solution of the system of the equations (\ref{acdiffeqfin}) - (\ref{b2b2diffeqfin}) and of the equation (\ref{diffsystcond}), instead of the general. Therefore, we take: 
\beq
\label{acsoluts}
a\left({\bar x}\right)=
{1\over{4i}}\;
{\rm sgn}\left(x^{-}\right)
\delta^{(2)}\left({\bf x}_{\perp}\right), 
\;\;\;\;\;\;\;\;\;\;\;\;  
c\left({\bar x}\right)=
-{1\over{4\sqrt{2}i}}
\mid\!{x^{-}}\!\mid
\left(\Delta_{\perp}-M^{2}\right)
\delta^{(2)}\left({\bf x}_{\perp}\right), 
\eeq
\beq
\label{d1d2soluts}
d_{1}\left({\bar x}\right)=
{1\over{4\alpha}}
\mid\!{x^{-}}\!\mid
\delta^{(2)}\left({\bf x}_{\perp}\right), 
\;\;\;\;\;\;\;\;\;\;\;\;   
d_{2}\left({\bar x}\right)=
-{1\over{4\alpha}}
\mid\!{x^{-}}\!\mid
\delta^{(2)}\left({\bf x}_{\perp}\right),
\eeq
\beq
\label{b1b2soluts}
b_{1}\left({\bar x}\right)=
-{1\over{8\alpha}}\;
{\rm sgn}\left(x^{-}\right)
\left(x^{-}\right)^{2}
\left(\Delta_{\perp}+{1\over{\alpha}}\right)
\delta^{(2)}\left({\bf x}_{\perp}\right), 
\;\;\;\;\;\;\;\;\;\;\;\;   
b_{2}\left({\bar x}\right)=
x^{-}B\left({\bf x}_{\perp}\right),   
\eeq
wherein the function $B({\bf x}_{\perp})$ obviously satisfies the same condition (\ref{Bsolut}). 

Here it is the free inverse Dirac-Bergmann matrix for the Yukawa model with higher order derivatives at our disposal, so that we can compute this matrix in the case of the interactions being switched on. 

\section{Inverse Dirac-Bergmann Matrix with Interactions}
\label{ifmi}

As mentioned earlier, we employ the first approach to obtain the inverse Dirac-Bergmann matrix with interactions. This is based on the series expansion (\ref{finverseperturbative0}), wherein the sequent terms within this pattern are described by the formulae   
(\ref{finverseperturbativen0}). In turn, the matrix containing the structure of the interactions is determined by the expression 
(\ref{gmatrixelements}). We are using the approach especially fit for that case, in which we meet the finite expansion 
(\ref{finverseperturbative0}). We commence the calculations from the three lowest order terms of the expansion of the inverse Dirac-Bergmann matrix with interactions ${\cal F}^{-1}={\cal F}^{-1}_{0}+{\cal A}_{1}+{\cal A}_{2}+{\cal A}_{3}$. The power of the truncation for the studied series is not accidental, as it is clear after the algebraic computations. According to the prescription (\ref{finverseperturbativen0}), we have: 
\beq
\label{expfmionr1}
{\cal A}_{1}\!\left(x^{+},{\bar x},{\bar y}\right)=
-\left\{
\int_{R^{3}}\!{d^{3}{\bar z}_{1}}\!
\int_{R^{3}}\!{d^{3}{\bar z}_{2}}\!\;
{\cal F}^{-1}_{0}\!\left(x^{+},{\bar z}_{1}\right)
{\cal G}\!\left(x^{+},{\bar z}_{1},{\bar z}_{2}\right)
{\cal F}^{-1}_{0}\!\left(x^{+},{\bar z}_{2},{\bar y}\right)
\right\},
\eeq
\beq
\label{expfmionr2}
{\cal A}_{2}\!\left(x^{+},{\bar x},{\bar y}\right)=
\eeq
\beqs
=\left\{
\int_{R^{3}}\!{d^{3}{\bar z}_{1}}\dots\!
\int_{R^{3}}\!{d^{3}{\bar z}_{4}}\!\;
{\cal F}^{-1}_{0}\left(x^{+},{\bar x},{\bar z}_{1}\right)
{\cal G}\!\left(x^{+},{\bar z}_{1},{\bar z}_{2}\right)
{\cal F}^{-1}_{0}\!\left(x^{+},{\bar z}_{2},{\bar z}_{3}\right)
{\cal G}\!\left(x^{+},{\bar z}_{3},{\bar z}_{4}\right)
{\cal F}^{-1}_{0}\!\left(x^{+},{\bar z}_{4},{\bar y}\right)
\right\},
\eeqs
\beq
\label{expfmionr3}
{\cal A}_{3}\!\left(x^{+},{\bar x},{\bar y}\right)=
\eeq
\beqs
=-\left\{
\int_{R^{3}}\!{d^{3}{\bar z}_{1}}\dots\!
\int_{R^{3}}\!{d^{3}{\bar z}_{6}}\!\;
{\cal F}^{-1}_{0}\!\left(x^{+},{\bar x},{\bar z}_{1}\right)
{\cal G}\!\left(x^{+},{\bar z}_{1},{\bar z}_{2}\right)
{\cal F}^{-1}_{0}\!\left(x^{+},{\bar z}_{2},{\bar z}_{3}\right)
{\cal G}\!\left(x^{+},{\bar z}_{3},{\bar z}_{4}\right)
\times
\right.
\eeqs
\beqs
\left.
\times
{\cal F}^{-1}_{0}\!\left(x^{+},{\bar z}_{4},{\bar z}_{5}\right)
{\cal G}\!\left(x^{+},{\bar z}_{5},{\bar z}_{6}\right)
{\cal F}^{-1}_{0}\!\left(x^{+},{\bar z}_{6},{\bar y}\right)
\right\}.
\eeqs

The first order contribution to the inverse Dirac-Bergmann matrix, described by the effective pattern (\ref{expfmionr1}), gives the final result after the insertion of the free inverse (\ref{invfmatrixelements}) - (\ref{invfmatrixelements5}) and of the interaction arrays (\ref{gmatrixelements}) - (\ref{gmatrix3}). It is easy to convince, that the algebraic and the analytical computations lead to 
\beq
\label{invfinr1} 
{\cal A}_{1}\!\left(x^{+},{\bar x},{\bar y}\right)=
-\left(
\begin{array}{cccccc}
0&\;\;
{\cal A}_{112}\left(x^{+},{\bar x},{\bar y}\right)&\;\;
{\cal A}_{113}\left(x^{+},{\bar x},{\bar y}\right)&\;\;
{\cal A}_{114}\left(x^{+},{\bar x},{\bar y}\right)&\;\;
0&\;\;0\\
{\cal A}_{121}\left(x^{+},{\bar x},{\bar y}\right)&\;\;
0&\;\;
{\cal A}_{123}\left(x^{+},{\bar x},{\bar y}\right)&\;\;
{\cal A}_{124}\left(x^{+},{\bar x},{\bar y}\right)&\;\; 
0&\;\;0\\
{\cal A}_{131}\left(x^{+},{\bar x},{\bar y}\right)&\;\;
{\cal A}_{132}\left(x^{+},{\bar x},{\bar y}\right)&\;\;
0&\;\;0&\;\;0&\;\;0\\
{\cal A}_{141}\left(x^{+},{\bar x},{\bar y}\right)&\;\;
{\cal A}_{142}\left(x^{+},{\bar x},{\bar y}\right)&\;\;
0&\;\;0&\;\;0&\;\;0\\
0&\;\;0&\;\;0&\;\;0&\;\;0&\;\;0\\ 
0&\;\;0&\;\;0&\;\;0&\;\;0&\;\;0\\
\end{array}
\right). 
\eeq
The elements of this array include the fermionic fields, the scalar ones and the functions derived from the inverse of the free Dirac-Bergmann matrix: $a(x^{+},{\bar x})$, $b_{2}(x^{+},{\bar x})$, $d_{1,2}(x^{+},{\bar x})$. They embrace mixed powers of the Yukawa coupling constant, as the method used for this considerations permits: 
\beq
\label{matra112}
{\cal A}_{112}\left(x^{+},{\bar x},{\bar y}\right)=
gF_{1}\left(x^{+},{\bar x},{\bar y}\right)+
g^{2}F^{2}_{1}\left(x^{+},{\bar x},{\bar y}\right),
\eeq
\beq
\label{matra113114}
{\cal A}_{113}\left(x^{+},{\bar x},{\bar y}\right)=
gF_{2}\left(x^{+},{\bar x},{\bar y}\right), 
\;\;\;\;\;\;\;\;\;\;\;\;   
{\cal A}_{114}\left(x^{+},{\bar x},{\bar y}\right)=
gF_{3}\left(x^{+},{\bar x},{\bar y}\right),
\eeq
\beq
\label{matra121}
{\cal A}_{121}\left(x^{+},{\bar x},{\bar y}\right)=
-gF_{1}\left(x^{+},{\bar x},{\bar y}\right)-
g^{2}F^{2}_{1}\left(x^{+},{\bar x},{\bar y}\right),
\eeq
\beq
\label{matra123124}
{\cal A}_{123}\left(x^{+},{\bar x},{\bar y}\right)=
gF_{4}\left(x^{+},{\bar x},{\bar y}\right), 
\;\;\;\;\;\;\;\;\;\;\;\;  
{\cal A}_{124}\left(x^{+},{\bar x},{\bar y}\right)=
gF_{5}\left(x^{+},{\bar x},{\bar y}\right),
\eeq
\beq
\label{matra131132}
{\cal A}_{131}\left(x^{+},{\bar x},{\bar y}\right)=
-gF_{6}\left(x^{+},{\bar x},{\bar y}\right), 
\;\;\;\;\;\;\;\;\;\;\;\;  
{\cal A}_{132}\left(x^{+},{\bar x},{\bar y}\right)=
-gF_{7}\left(x^{+},{\bar x},{\bar y}\right),
\eeq
\beq
\label{matra141}
{\cal A}_{141}\left(x^{+},{\bar x},{\bar y}\right)=
-gF_{8}\left(x^{+},{\bar x},{\bar y}\right), 
\;\;\;\;\;\;\;\;\;\;\;\;   
{\cal A}_{142}\left(x^{+},{\bar x},{\bar y}\right)=
-gF_{9}\left(x^{+},{\bar x},{\bar y}\right). 
\eeq
Moreover, we have the following description of these functions. Their specific expressions contain the field operators in the linear fermionic and also in the quadratic scalar terms: 
\beq
\label{invfinr1elements1}
F_{1}\left(x^{+},{\bar x},{\bar y}\right)=
2\sqrt{2}iM\!\!
\int_{R^{3}}\!{d^{3}{\bar z}}\;
a\left(x^{+},{\bar x}-{\bar z}\right)
{\phi}\left(x^{+},{\bar z}\right)
a\left(x^{+},{\bar z}-{\bar y}\right)
\Lambda_{-},
\eeq
\beq
\label{invfinr1elements11}
F^{2}_{1}\left(x^{+},{\bar x},{\bar y}\right)=
\sqrt{2}i\!\!
\int_{R^{3}}\!{d^{3}{\bar z}}\;
a\left(x^{+},{\bar x}-{\bar z}\right)
{\phi}^{2}\left(x^{+},{\bar z}\right)
a\left(x^{+},{\bar z}-{\bar y}\right)
\Lambda_{-},
\eeq
\beq
\label{invfinr1elements2}
F_{2}\left(x^{+},{\bar x},{\bar y}\right)=
-\!\int_{R^{3}}\!{d^{3}{\bar z}}\;
a\left(x^{+},{\bar x}-{\bar z}\right)
\gamma^{+}\Psi_{+}\left(x^{+},{\bar z}\right)
d_{2}\left(x^{+},{\bar z}-{\bar y}\right),
\eeq 
\beq
\label{invfinr1elements3}
F_{3}\left(x^{+},{\bar x},{\bar y}\right)=
-\int_{R^{3}}\!{d^{3}{\bar z}}\;
a\left(x^{+},{\bar x}-{\bar z}\right)
\gamma^{+}\Psi_{+}\left(x^{+},{\bar z}\right)
b_{2}\left(x^{+},{\bar z}-{\bar y}\right),
\eeq
\beq
\label{invfinr1elements4}
F_{4}\left(x^{+},{\bar x},{\bar y}\right)=
\int_{R^{3}}\!{d^{3}{\bar z}}\;
a\left(x^{+},{\bar x}-{\bar z}\right)
\Psi^{\dagger}_{+}\left(x^{+},{\bar z}\right)\gamma^{-}
d_{2}\left(x^{+},{\bar z}-{\bar y}\right),
\eeq
\beq
\label{invfinr1elements5}
F_{5}\left(x^{+},{\bar x},{\bar y}\right)=
\!\int_{R^{3}}\!{d^{3}{\bar z}}\;
a\left(x^{+},{\bar x}-{\bar z}\right)
\Psi^{\dagger}_{+}\left(x^{+},{\bar z}\right)\gamma^{-}
b_{2}\left(x^{+},{\bar z}-{\bar y}\right),
\eeq
\beq
\label{invfinr1elements6}
F_{6}\left(x^{+},{\bar x},{\bar y}\right)=
\!\int_{R^{3}}\!{d^{3}{\bar z}}\;
d_{1}\left(x^{+},{\bar x}-{\bar z}\right)
\gamma^{+}\Psi_{+}\left(x^{+},{\bar z}\right)
a\left(x^{+},{\bar z}-{\bar y}\right),
\eeq
\beq
\label{invfinr1elements7}
F_{7}\left(x^{+},{\bar x},{\bar y}\right)=
-\!\int_{R^{3}}\!{d^{3}{\bar z}}\;
d_{1}\left(x^{+},{\bar x}-{\bar z}\right)
\Psi^{\dagger}_{+}\left(x^{+},{\bar z}\right)\gamma^{-}
a\left(x^{+},{\bar z}-{\bar y}\right),
\eeq
\beq
\label{invfinr1elements8}
F_{8}\left(x^{+},{\bar x},{\bar y}\right)=
\!\int_{R^{3}}\!{d^{3}{\bar z}}\;
b_{2}\left(x^{+},{\bar x}-{\bar z}\right)
\gamma^{+}\Psi_{+}\left(x^{+},{\bar z}\right)
a\left(x^{+},{\bar z}-{\bar y}\right),
\eeq
\beq
\label{invfinr1elements9}
F_{9}\left(x^{+},{\bar x},{\bar y}\right)=
-\!\int_{R^{3}}\!{d^{3}{\bar z}}\;
b_{2}\left(x^{+},{\bar x}-{\bar z}\right)
\Psi^{\dagger}_{+}\left(x^{+},{\bar z}\right)\gamma^{-}
a\left(x^{+},{\bar z}-{\bar y}\right). 
\eeq

The second order contribution ${\cal A}_{2}$ is described by the formula analogous to the ${\cal A}_{1}$, but includes longer chain of the 
${\cal F}^{-1}_{0}$ and of the ${\cal G}$ matrices (\ref{expfmionr2}). The algebraic and the analytical computations allow us to put final, effective expression for this contribution as the array 
\beq
\label{invfinr2}
{\cal A}_{2}\left(x^{+},{\bar x},{\bar y}\right)=
\left(
\begin{array}{cccccc}
{\cal A}_{211}\left(x^{+},{\bar x},{\bar y}\right)&\;\;
{\cal A}_{212}\left(x^{+},{\bar x},{\bar y}\right)&\;\;
0&\;\;0&\;\;0&\;\;0\\
{\cal A}_{221}\left(x^{+},{\bar x},{\bar y}\right)&\;\;
{\cal A}_{222}\left(x^{+},{\bar x},{\bar y}\right)&\;\;
0&\;\;0&\;\;0&\;\;0\\
0&\;\;0&\;\;0&\;\;0&\;\;0&\;\;0\\
0&\;\;0&\;\;0&\;\;0&\;\;0&\;\;0\\ 
0&\;\;0&\;\;0&\;\;0&\;\;0&\;\;0\\
0&\;\;0&\;\;0&\;\;0&\;\;0&\;\;0\\
\end{array}
\right),
\eeq
which embraces the integrals consisting of the function $a(x^{+},{\bar x})$ and of the $b_{2}(x^{+},{\bar x})$, taken from the inverse of the free Dirac-Bergmann matrix. They are proportional to the second power of the Yukawa coupling constant. It means, that: 
\beq
\label{invfinr2elements12}
{\cal A}_{211}\left(x^{+},{\bar x},{\bar y}\right)=
g^{2}H_{1}\left(x^{+},{\bar x},{\bar y}\right), \;\;\;\;\;\; 
{\cal A}_{212}\left(x^{+},{\bar x},{\bar y}\right)=
-g^{2}H_{2}\left(x^{+},{\bar x},{\bar y}\right),
\eeq
\beq
\label{invfinr2elements34}
{\cal A}_{221}\left(x^{+},{\bar x},{\bar y}\right)=
g^{2}H_{3}\left(x^{+},{\bar x},{\bar y}\right), \;\;\;\;\;\; 
{\cal A}_{222}\left(x^{+},{\bar x},{\bar y}\right)=
-g^{2}H_{4}\left(x^{+},{\bar x},{\bar y}\right).   
\eeq
The detailed structure of the above functions reveals the bi-linear tensor products of the fermionic fields: 
\beq
\label{invhfunct1}
H_{1}\left(x^{+},{\bar x},{\bar y}\right)=
\eeq
\beqs
=\!\int_{R^{3}}\!{d^{3}{\bar z}_{1}}\!
\int_{R^{3}}\!{d^{3}{\bar z}_{2}}\;
a\left(x^{+},{\bar x}-{\bar z}_{1}\right)
\gamma^{+}{\Psi}_{+}\left(x^{+},{\bar z}_{1}\right)
b_{2}\left(x^{+},{\bar z}_{1}-{\bar z}_{2}\right)
\gamma^{+}{\Psi}_{+}\left(x^{+},{\bar z}_{2}\right)
a\left(x^{+},{\bar z}_{2}-{\bar y}\right),
\eeqs
\beq
\label{invhfunct2}
H_{2}\left(x^{+},{\bar x},{\bar y}\right)=
\eeq
\beqs
=\int_{R^{3}}\!{d^{3}{\bar z}_{1}}\!
\int_{R^{3}}\!{d^{3}{\bar z}_{2}}\;
a\left(x^{+},{\bar x}-{\bar z}_{1}\right)
\gamma^{+}{\Psi}_{+}\left(x^{+},{\bar z}_{1}\right)
b_{2}\left(x^{+},{\bar z}_{1}-{\bar z}_{2}\right)
{\Psi}^{\dagger}_{+}\left(x^{+},{\bar z}_{2}\right)\gamma^{-}
a\left(x^{+},{\bar z}_{2}-{\bar y}\right),
\eeqs 
\beq
\label{invhfunct3}
H_{3}\left(x^{+},{\bar x},{\bar y}\right)=
\eeq
\beqs
=\int_{R^{3}}\!{d^{3}{\bar z}_{1}}\!
\int_{R^{3}}\!{d^{3}{\bar z}_{2}}\;
a\left(x^{+},{\bar x}-{\bar z}_{1}\right)
{\Psi}^{\dagger}_{+}\left(x^{+},{\bar z}_{1}\right)\gamma^{-}
b_{2}\left(x^{+},{\bar z}_{1}-{\bar z}_{2}\right)
\gamma^{+}{\Psi}_{+}\left(x^{+},{\bar z}_{2}\right)
a\left(x^{+},{\bar z}_{2}-{\bar y}\right),
\eeqs 
\beq
\label{invhfunct4}
H_{4}\left(x^{+},{\bar x},{\bar y}\right)=
\eeq
\beqs
=-\!\int_{R^{3}}\!{d^{3}{\bar z}_{1}}\!
\int_{R^{3}}\!{d^{3}{\bar z}_{2}}\;
a\left(x^{+},{\bar x}-{\bar z}_{1}\right)
{\Psi}^{\dagger}_{+}\left(x^{+},{\bar z}_{1}\right)\gamma^{-}
b_{2}\left(x^{+},{\bar z}_{1}-{\bar z}_{2}\right)
{\Psi}^{\dagger}_{+}\left(x^{+},{\bar z}_{2}\right)\gamma^{-}
a\left(x^{+},{\bar z}_{2}-{\bar y}\right).
\eeqs

The form of the third order contribution ${\cal A}_{3}$ to the inverse Dirac-Bergmann matrix with interactions is determined by the pattern (\ref{expfmionr3}) and includes the pure triple product of the arrays, denoted in the parentheses. This is just trivial 
$({\cal F}^{-1}_{0}{\cal G})({\cal F}^{-1}_{0}{\cal G})({\cal F}^{-1}_{0}{\cal G})=0$. Thus, the third order contribution to the inverse Dirac-Bergmann matrix satisfies ${\cal A}_{3}=0$. And furthermore, the general prescription (\ref{finverseperturbativen0}) unveils its iterative feature, wherein the $n$th order of the contribution is algebraically determined by the direct predecessor of the $(n-1)$th order, then multiplied by the matrix chain link $({\cal F}^{-1}_{0}{\cal G})$. Therefore, all the third and the higher order contributions to the inverse Dirac-Bergmann matrix for the interacting Yukawa model with Bernard-Duncan term vanish ${\cal A}_{n}(n\ge3)=0$.

It can be seen, that the final array ${\cal F}^{-1}$ is determined by the truncated and effective expansion 
${\cal F}^{-1}={\cal F}^{-1}_{0}+{\cal A}_{1}+{\cal A}_{2}$. This depends: on the functions (\ref{asolut}) - (\ref{b2solut}) within the free matrix ${\cal F}^{-1}_{0}$, on the first (\ref{invfinr1elements1}) - (\ref{invfinr1elements9}) and also on the second (\ref{invhfunct1}) - (\ref{invhfunct4}) order contributions. In view of the above, we can conclude, that the end result for the matrix 
${\cal F}^{-1}(x^{+},{\bar x},{\bar y})$ in discussed case is entirely finite 
\beq
\label{finvfin}
{\cal F}^{-1}\!\left(x^{+},{\bar x},{\bar y}\right)=
\eeq
\beqs
=\left(
\begin{array}{cccccc}
{\cal F}^{-1}_{{11}_{{\mbox{ }}_{{\mbox{ }}_{\mbox{ }}}}}\!\!\!\!\!\!\left(x^{+},{\bar x},{\bar y}\right)&\;\;\;
{\cal F}^{-1}_{12}\!\left(x^{+},{\bar x},{\bar y}\right)&\;\;\;
{\cal F}^{-1}_{13}\!\left(x^{+},{\bar x},{\bar y}\right)&\;\;\;
{\cal F}^{-1}_{14}\!\left(x^{+},{\bar x},{\bar y}\right)&\;\;\;
0&\;\;\;
{\cal F}^{-1}_{16}\!\left(x^{+},{\bar x},{\bar y}\right)\\
{\cal F}^{-1}_{{21}_{{\mbox{ }}_{{\mbox{ }}_{\mbox{ }}}}}\!\!\!\!\!\!\left(x^{+},{\bar x},{\bar y}\right)&\;\;\;
{\cal F}^{-1}_{22}\!\left(x^{+},{\bar x},{\bar y}\right)&\;\;\;
{\cal F}^{-1}_{23}\!\left(x^{+},{\bar x},{\bar y}\right)&\;\;\;
{\cal F}^{-1}_{24}\!\left(x^{+},{\bar x},{\bar y}\right)&\;\;\;
{\cal F}^{-1}_{25}\!\left(x^{+},{\bar x},{\bar y}\right)&\;\;\;
0\\
{\cal F}^{-1}_{{31}_{{\mbox{ }}_{{\mbox{ }}_{\mbox{ }}}}}\!\!\!\!\!\!\left(x^{+},{\bar x},{\bar y}\right)&\;\;\;
{\cal F}^{-1}_{32}\!\left(x^{+},{\bar x},{\bar y}\right)&\;\;\; 
{\cal F}^{-1}_{33}\!\left(x^{+},{\bar x},{\bar y}\right)&\;\;\;
{\cal F}^{-1}_{34}\!\left(x^{+},{\bar x},{\bar y}\right)&\;\;\;   
0&\;\;\;0\\
{\cal F}^{-1}_{{41}_{{\mbox{ }}_{{\mbox{ }}_{\mbox{ }}}}}\!\!\!\!\!\!\left(x^{+},{\bar x},{\bar y}\right)&\;\;\;
{\cal F}^{-1}_{42}\!\left(x^{+},{\bar x},{\bar y}\right)&\;\;\; 
{\cal F}^{-1}_{43}\!\left(x^{+},{\bar x},{\bar y}\right)&\;\;\;
{\cal F}^{-1}_{44}\!\left(x^{+},{\bar x},{\bar y}\right)&\;\;\;   
0&\;\;\;0\\
\;\;\;\;0_{{ }_{{\mbox{ }}_{{\mbox{ }}_{\mbox{ }}}}}&\;\;\;
{\cal F}^{-1}_{52}\!\left(x^{+},{\bar x},{\bar y}\right)&\;\;\; 
0&\;\;\;0&\;\;\;0&\;\;\;0\\ 
{\cal F}^{-1}_{61}\!\left(x^{+},{\bar x},{\bar y}\right)&\;\;\; 
0&\;\;\;0&\;\;\;0&\;\;\;0&\;\;\;0\\
\end{array}
\right)\!\!. 
\eeqs
The elements of this array are displayed below:   
\beq
\label{finvfinelem11}
{\cal F}^{-1}_{11}\left(x^{+},{\bar x},{\bar y}\right)=
g^{2}H_{1}\left(x^{+},{\bar x},{\bar y}\right),
\eeq
\beq
\label{finvfinelem12} 
{\cal F}^{-1}_{12}\left(x^{+},{\bar x},{\bar y}\right)=
c\left(x^{+},{\bar x}-{\bar y}\right)\Lambda_{-}+
gF_{1}\left(x^{+},{\bar x},{\bar y}\right)+
g^{2}F^{2}_{1}\left(x^{+},{\bar x},{\bar y}\right)-
g^{2}H_{2}\left(x^{+},{\bar x},{\bar y}\right), 
\eeq
\beq
\label{finvfinelem1314}
{\cal F}^{-1}_{13}\left(x^{+},{\bar x},{\bar y}\right)=
gF_{2}\left(x^{+},{\bar x},{\bar y}\right), 
\;\;\;\;\;\;\;\;\;\;\;\;   
{\cal F}^{-1}_{14}\left(x^{+},{\bar x},{\bar y}\right)=
gF_{3}\left(x^{+},{\bar x},{\bar y}\right), 
\eeq
\beq
\label{finvfinelem16} 
{\cal F}^{-1}_{16}\left(x^{+},{\bar x},{\bar y}\right)=
-a\left(x^{+},{\bar x}-{\bar y}\right)\Lambda_{-}, 
\eeq
\beq
\label{finvfinelem21}
{\cal F}^{-1}_{21}\left(x^{+},{\bar x},{\bar y}\right)=
-c\left(x^{+},{\bar x}-{\bar y}\right)\Lambda_{-}-
gF_{1}\left(x^{+},{\bar x},{\bar y}\right)-
g^{2}F^{2}_{1}\left(x^{+},{\bar x},{\bar y}\right)+
g^{2}H_{3}\left(x^{+},{\bar x},{\bar y}\right),
\eeq
\beq
\label{finvfinelem2223}
{\cal F}^{-1}_{22}\left(x^{+},{\bar x},{\bar y}\right)=
-g^{2}H_{4}\left(x^{+},{\bar x},{\bar y}\right), 
\;\;\;\;\;\;\;\;\;\;\;\;   
{\cal F}^{-1}_{23}\left(x^{+},{\bar x},{\bar y}\right)=
gF_{4}\left(x^{+},{\bar x},{\bar y}\right), 
\eeq 
\beq
\label{finvfinelem2425} 
{\cal F}^{-1}_{24}\left(x^{+},{\bar x},{\bar y}\right)=
gF_{5}\left(x^{+},{\bar x},{\bar y}\right), 
\;\;\;\;\;\;\;\;\;\;\;\;  
{\cal F}^{-1}_{25}\left(x^{+},{\bar x},{\bar y}\right)=
-a\left(x^{+},{\bar x}-{\bar y}\right)\Lambda_{-},  
\eeq
\beq
\label{finvfinelem3132} 
{\cal F}^{-1}_{31}\left(x^{+},{\bar x},{\bar y}\right)=
-gF_{6}\left(x^{+},{\bar x},{\bar y}\right), 
\;\;\;\;\;\;\;\;\;\;\;\;  
{\cal F}^{-1}_{32}\left(x^{+},{\bar x},{\bar y}\right)=
-gF_{7}\left(x^{+},{\bar x},{\bar y}\right),  
\eeq
\beq
\label{finvfinelem3334} 
{\cal F}^{-1}_{33}\left(x^{+},{\bar x},{\bar y}\right)=
b_{1}\left(x^{+},{\bar x}-{\bar y}\right){\rm P}, 
\;\;\;\;\;\;\;\;\;\;\;\;  
{\cal F}^{-1}_{34}\left(x^{+},{\bar x},{\bar y}\right)=
d_{1}\left(x^{+},{\bar x}-{\bar y}\right){\rm P},  
\eeq
\beq
\label{finvfinelem4142} 
{\cal F}^{-1}_{41}\left(x^{+},{\bar x},{\bar y}\right)=
-gF_{8}\left(x^{+},{\bar x},{\bar y}\right), 
\;\;\;\;\;\;\;\;\;\;\;\;  
{\cal F}^{-1}_{42}\left(x^{+},{\bar x},{\bar y}\right)=
-gF_{9}\left(x^{+},{\bar x},{\bar y}\right),  
\eeq
\beq
\label{finvfinelem4344} 
{\cal F}^{-1}_{43}\left(x^{+},{\bar x},{\bar y}\right)=
d_{2}\left(x^{+},{\bar x}-{\bar y}\right){\rm P}, 
\;\;\;\;\;\;\;\;\;\;\;\;  
{\cal F}^{-1}_{44}\left(x^{+},{\bar x},{\bar y}\right)=
b_{2}\left(x^{+},{\bar x}-{\bar y}\right){\rm P},  
\eeq
\beq
\label{finvfinelem5261} 
{\cal F}^{-1}_{52}\left(x^{+},{\bar x},{\bar y}\right)=
a\left(x^{+},{\bar x}-{\bar y}\right)\Lambda_{-}, 
\;\;\;\;\;\;\;\;\;\;\;\;  
{\cal F}^{-1}_{61}\left(x^{+},{\bar x},{\bar y}\right)=
a\left(x^{+},{\bar x}-{\bar y}\right)\Lambda_{-}.  
\eeq

The inverse Dirac-Bergmann matrix ${\cal F}^{-1}(x^{+},{\bar x},{\bar y})$, described through the above functions (\ref{finvfinelem11}) - (\ref{finvfinelem5261}), incorporates the information about the interactions within analyzed model, originating in the Lagrange equations of motion. From this point of view presented method gives the exact solution of the studied problem. But this observation refers only to the case of the finite series (\ref{finverseperturbative0}), like in the Yukawa model. When formulating it in some other way, there the question rises, whether already introduced method is only perturbative and whether the exact. Of course, it also depends on the complexity of the studied model (solvable or not), but this does not exhaust the topics, especially for the functional methods, to which our approach belongs \cite{b22,b23,b24,b25}. 

\section{(Anti-) commutators for Interacting Fields}
\label{dbym}

The final results of our method (\ref{finvfin}) - (\ref{finvfinelem5261}), devoted to obtaining the inverse matrix of the constraints for the interacting Yukawa model, may be directly applied to compute the Dirac brackets from the Poisson ones, according to the prescription (\ref{dirbracketdef}). We take into considerations the canonical Poisson brackets for our model: (\ref{canquant1}), (\ref{canquant3}), the equations of the constraints: (\ref{constraintsprimary}), (\ref{constraintprimary0}), (\ref{constraintsprimary1}), (\ref{constraintssecondary4}), (\ref{constraintssecondary5}) and we insert all these expressions into the pattern (\ref{dirbracketdef}). Next, we apply the procedure of the quantization, carrying out the replacement $\{\cdot,\cdot\}_{\!\mbox{}_{D,P}}\rightarrow(1/i)[\cdot,\cdot]$ or 
$\{\cdot,\cdot\}_{\!\mbox{}_{D,P}}\rightarrow(1/i)\{\cdot,\cdot\}$, due to the gradation of the operator algebra \cite{b26,b27}. Thus: 
\beq
\label{dirbracketpsippsipf}
\left\{
\Psi_{+}\left(x^{+},{\bar x}\right) \; , \; \Psi_{+}\left(x^{+},{\bar y}\right)
\right\}=
\left\{
\Psi_{+}\left(x^{+},{\bar x}\right) \; , \; \Psi_{+}\left(x^{+},{\bar y}\right)
\right\}_{\rm{can.}}
=0, 
\eeq
\beq
\label{dirbracketpsippsidpf}
\left\{
\Psi_{+}\left(x^{+},{\bar x}\right) \; , \; \Psi^{\dagger}_{+}\left(x^{+},{\bar y}\right)
\right\}=
\left\{
\Psi_{+}\left(x^{+},{\bar x}\right) \; , \; \Psi^{\dagger}_{+}\left(x^{+},{\bar y}\right)
\right\}_{\rm{can.}}=
{1\over{\sqrt{2}}}
\delta^{(3)}\left({\bar x}-{\bar y}\right)
\Lambda_{+},
\eeq
\beq
\label{dirbracketpsimpsipf}
\left\{
\Psi_{+}\left(x^{+},{\bar x}\right) \; , \; \Psi_{-}\left(x^{+},{\bar y}\right)
\right\}=
\left\{
\Psi_{+}\left(x^{+},{\bar x}\right) \; , \; \Psi_{-}\left(x^{+},{\bar y}\right)
\right\}_{\rm{can.}}
=0, 
\eeq
\beq
\label{dirbracketpsimpsidpf}
\left\{
\Psi_{+}\left(x^{+},{\bar x}\right) \; , \; \Psi^{\dagger}_{-}\left(x^{+},{\bar y}\right)
\right\}=
{\underbrace{
\left\{
\Psi_{+}\left(x^{+},{\bar x}\right) \; , \; \Psi^{\dagger}_{-}\left(x^{+},{\bar y}\right)
\right\}_{\rm{can.}}
}_{0}}-
\eeq
\beqs
-{1\over{\sqrt{2}}}
\left\{
i\!\not\!\partial^{x}_{\perp}a\left(x^{+},{\bar x}-{\bar y}\right)+
a\left(x^{+},{\bar x}-{\bar y}\right)
\left[M+g\phi\left(x^{+},{\bar x}\right)\right]
\right\}
\gamma^{-},
\eeqs
\beq
\label{dirbracketpsimpsimf}
\left\{
\Psi_{-}\left(x^{+},{\bar x}\right) \; , \; \Psi_{-}\left(x^{+},{\bar y}\right)
\right\}=
{\underbrace{
\left\{
\Psi_{-}\left(x^{+},{\bar x}\right) \; , \; \Psi_{-}\left(x^{+},{\bar y}\right)
\right\}_{\rm{can.}}
}_{0}}+
ig^{2}H_{1}\left(x^{+},{\bar x},{\bar y}\right), 
\eeq
\beq
\label{dirbracketpsimpsidmf}
\left\{
\Psi_{-}\left(x^{+},{\bar x}\right) \; , \; \Psi^{\dagger}_{-}\left(x^{+},{\bar y}\right)
\right\}=
{\underbrace{
\left\{
\Psi_{-}\left(x^{+},{\bar x}\right) \; , \; \Psi^{\dagger}_{-}\left(x^{+},{\bar y}\right)
\right\}_{\rm{can.}}
}_{0}}+
\eeq
\beqs
+ic\left(x^{+},{\bar x}-{\bar y}\right)\Lambda_{-}+
igF_{1}\left(x^{+},{\bar x},{\bar y}\right)+
ig^{2}F^{2}_{1}\left(x^{+},{\bar x},{\bar y}\right)-
ig^{2}H_{2}\left(x^{+},{\bar x},{\bar y}\right),
\eeqs
\beq
\label{dirbracketphipsipf}
\left[
\Psi_{+}\left(x^{+},{\bar x}\right) \; , \; \phi\left(x^{+},{\bar y}\right) 
\right]=
\left[
\Psi_{+}\left(x^{+},{\bar x}\right) \; , \, \phi\left(x^{+},{\bar y}\right) 
\right]_{\rm{can.}}
=0, 
\eeq
\beq
\label{dirbracketpsippartpphi}
\left[
\Psi_{+}\left(x^{+},{\bar x}\right) \; , \; \partial_{+}\phi\left(x^{+},{\bar y}\right)
\right]=
\left[
\Psi_{+}\left(x^{+},{\bar x}\right) \; , \; \partial_{+}\phi\left(x^{+},{\bar y}\right)
\right]_{\rm{can.}}
=0,
\eeq
\beq
\label{dirbracketphipsimf}
\left[
\Psi_{-}\left(x^{+},{\bar x}\right) \; , \; \phi\left(x^{+},{\bar y}\right)
\right]=
{\underbrace{
\left[
\Psi_{-}\left(x^{+},{\bar x}\right) \; , \; \phi\left(x^{+},{\bar y}\right)
\right]_{\rm{can.}}
}_{0}}+
igF_{3}\left(x^{+},{\bar x},{\bar y}\right),
\eeq
\beq
\label{dirbracketpartpphipsimf}
\left[
\Psi_{-}\left(x^{+},{\bar x}\right) \; , \; \partial_{+}\phi\left(x^{+},{\bar y}\right) 
\right]=
{\underbrace{
\left[
\Psi_{-}\left(x^{+},{\bar x}\right) \; , \; \partial_{+}\phi\left(x^{+},{\bar y}\right) 
\right]_{\rm{can.}}
}_{0}}+
igF_{6}\left(x^{+},{\bar y},{\bar x}\right),
\eeq
\beq
\label{dirbracketphiphif}
\left[
\phi\left(x^{+},{\bar x}\right) \; , \; \phi\left(x^{+},{\bar y}\right)
\right]=
{\underbrace{
\left[
\phi\left(x^{+},{\bar x}\right) \; , \; \phi\left(x^{+},{\bar y}\right)
\right]_{\rm{can.}}
}_{0}}+
ib_{2}\left(x^{+},{\bar x}-{\bar y}\right),
\eeq
\beq
\label{dirbracketphiparpphif}
\left[
\phi\left(x^{+},{\bar x}\right) \; , \; \partial_{+}\phi\left(x^{+},{\bar y}\right)
\right]=
{\underbrace{
\left[
\phi\left(x^{+},{\bar x}\right) \; , \; \partial_{+}\phi\left(x^{+},{\bar y}\right)
\right]_{\rm{can.}}
}_{0}}-
id_{1}\left(x^{+},{\bar y}-{\bar x}\right), 
\eeq
\beq
\label{dirbracketparpphiparpphif}
\left[
\partial_{+}\phi\left(x^{+},{\bar x}\right) \; , \; \partial_{+}\phi\left(x^{+},{\bar y}\right)
\right]=
{\underbrace{
\left[
\partial_{+}\phi\left(x^{+},{\bar x}\right) \; , \; \partial_{+}\phi\left(x^{+},{\bar y}\right)
\right]_{\rm{can.}}
}_{0}}+
ib_{1}\left(x^{+},{\bar x}-{\bar y}\right).
\eeq

The remaining part of this algebra may easily be obtained from the above expressions by the Hermitian conjugation, what leads to another, but not displayed here (anti-) commutators, which include one or two fermionic fields as their arguments. Such abbreviated way of dealing is allowed due to the fact, that presented method exploits all of the constraints and finally leads to the consistent set of the (anti-) commutators. The similar situation recurs for the derivatives of the discussed algebra with respect to the coordinates $x^{-}$ and to the $x^{j}$, where $j=1,2$. The fields: $\partial_{\bar{\mu}}\Psi_{\pm}$, $\partial_{\bar{\mu}}\Psi^{\dagger}_{\pm}$, $\partial_{\bar{\mu}}\phi$ for $\bar{\mu}=-,j$ are not fundamental in our approach and therefore, the (anti-) commutators with presence of them can by straightforwardly computed by the relevant differentiation. 

Let us demonstrate the structure of the above algebra in greater detail. The expressions: (\ref{dirbracketpsimpsimf}), (\ref{dirbracketpsimpsidmf}), (\ref{dirbracketphipsimf}), (\ref{dirbracketpartpphipsimf}) describe some of the elements of this algebra and include functions taken from the procedure of computing the inverse Dirac-Bergmann matrix: $F_{1,3,6}(x^{+},{\bar x},{\bar y})$, $F^{2}_{1}(x^{+},{\bar x},{\bar y})$, $H_{1,2}(x^{+},{\bar x},{\bar y})$. These functions are defined by: (\ref{invfinr1elements1}), (\ref{invfinr1elements3}), (\ref{invfinr1elements6}), (\ref{invfinr1elements11}), (\ref{invhfunct1}) and (\ref{invhfunct2}). Herein we have, under the integral sign of these objects, the elements of the free inverse Dirac-Bergmann matrix: $a(x^{+},{\bar x})$, $b_{2}(x^{+},{\bar x})$, $d_{1,2}(x^{+},{\bar x})$. They satisfy the differential equations in the domain of the distribution (\ref{acdiffeqfin}) - (\ref{b2b2diffeqfin}), which right sides are, among others, the Dirac deltas. The general solutions of these equations are (\ref{asolut}) - (\ref{b2solut}) and they allow us to bring relevant (anti-) commutators to the form of the local relationships, which consist the linear or the bi-linear combinations of the indefinite integrals of the fundamental fields. We introduced these integrals via the symbol $\partial^{-1}_{-}$ or the $\partial^{-2}_{-}$, for the single or the double indefinite integral of the field with respect to the $x^{-}$ variable. The integration by parts of: (\ref{invfinr1elements1}), (\ref{invfinr1elements11}), (\ref{invfinr1elements3}), (\ref{invfinr1elements6}), (\ref{invhfunct1}), (\ref{invhfunct2}) with respect to the variable $z^{-}$ reveals also, that occurring limit terms for $z^{-}\rightarrow\mp\infty$ may be singular and thus, the whole of the final expressions of this kind could demand regularization. Therefore, we can rewrite the anti-commutator (\ref{dirbracketpsimpsimf}) as 
\beq
\label{dirbracketpsimpsimfint}
\left\{
\Psi_{-}\left(x^{+},x^{-},{\bf x}_{\perp}\right) \; , \; \Psi_{-}\left(x^{+},y^{-},{\bf y}_{\perp}\right)
\right\}=
ig^{2}h^{\rm{reg.}}_{1}\left(x^{+},x^{-},{\bf x}_{\perp},y^{-},{\bf y}_{\perp}\right)-
\eeq
\beqs
-{i\over 4}g^{2}\gamma^{+}
\left(\partial^{-2}_{-}\Psi_{+}\right)\left(x^{+},x^{-},{\bf x}_{\perp}\right)
B\left({\bf x}_{\perp}-{\bf y}_{\perp}\right)
\gamma^{+}
\left(\partial^{-1}_{-}\Psi_{+}\right)\left(x^{+},y^{-},{\bf y}_{\perp}\right)+
\eeqs
\beqs
+{i\over 4}g^{2}\gamma^{+}
\left(\partial^{-1}_{-}\Psi_{+}\right)\left(x^{+},x^{-},{\bf x}_{\perp}\right)
B\left({\bf x}_{\perp}-{\bf y}_{\perp}\right)
\gamma^{+}
\left(\partial^{-2}_{-}\Psi_{+}\right)\left(x^{+},y^{-},{\bf y}_{\perp}\right)+ 
\eeqs
\beqs
+{i\over 4}g^{2}\gamma^{+} 
\left(\partial^{-1}_{-}\Psi_{+}\right)\left(x^{+},x^{-},{\bf x}_{\perp}\right)
\Big[
\left(x^{-}-y^{-}\right)B\left({\bf x}_{\perp}-{\bf y}_{\perp}\right)+
b_{20}\left({\bf x}_{\perp}-{\bf y}_{\perp}\right)
\Big]
\gamma^{+}
\left(\partial^{-1}_{-}\Psi_{+}\right)\left(x^{+},y^{-},{\bf y}_{\perp}\right). 
\eeqs
Similarly, the equation (\ref{dirbracketpsimpsidmf}) takes the form  
\beq
\label{dirbracketpsimpsimfint}
\left\{
\Psi_{-}\left(x^{+},x^{-},{\bf x}_{\perp}\right) \; , \; \Psi^{\dagger}_{-}\left(x^{+},y^{-},{\bf y}_{\perp}\right)
\right\}=
\eeq
\beqs
=igf^{\;\rm{reg.}}_{1}\left(x^{+},x^{-},{\bf x}_{\perp},y^{-},{\bf y}_{\perp}\right)+
ig^{2}f^{2\;\rm{reg.}}_{1}\left(x^{+},x^{-},{\bf x}_{\perp},y^{-},{\bf y}_{\perp}\right)-
ig^{2}h^{\rm{reg.}}_{2}\left(x^{+},x^{-},{\bf x}_{\perp},y^{-},{\bf y}_{\perp}\right)-
\eeqs
\beqs
-{i\over\sqrt{2}}
\left(x^{-}-y^{-}\right)\!
\left(\Delta^{x}_{\perp}-M^{2}\right)\!
\left[
{1\over{4i}}\;
{\rm{sgn}}\left(x^{-}-y^{-}\right)
\delta^{(2)}\left({\bf x}_{\perp}-{\bf y}_{\perp}\right)\!+\!
a_{0}\left({\bf x}_{\perp}-{\bf y}_{\perp}\right)
\right]\!
\Lambda_{-}+
ic_{0}\left({\bf x}_{\perp}-{\bf y}_{\perp}\right)\!
\Lambda_{-}+
\eeqs
\beqs
+\!{i\over 4}g^{2}\gamma^{+}
\left(\partial^{-2}_{-}\Psi_{+}\right)\left(x^{+},x^{-},{\bf x}_{\perp}\right)
B\left({\bf x}_{\perp}-{\bf y}_{\perp}\right)
\left(\partial^{-1}_{-}\Psi^{\dagger}_{+}\right)\left(x^{+},y^{-},{\bf y}_{\perp}\right)
\gamma^{-}-
\eeqs
\beqs
-{i\over 4}g^{2}\gamma^{+}
\left(\partial^{-1}_{-}\Psi_{+}\right)\left(x^{+},x^{-},{\bf x}_{\perp}\right)
B\left({\bf x}_{\perp}-{\bf y}_{\perp}\right)
\left(\partial^{-2}_{-}\Psi^{\dagger}_{+}\right)\left(x^{+},y^{-},{\bf y}_{\perp}\right)
\gamma^{-}-
\eeqs
\beqs
-{i\over 4}g^{2}
\gamma^{+}
\left(\partial^{-1}_{-}\Psi_{+}\right)\left(x^{+},x^{-},{\bf x}_{\perp}\right)
\left[
\left(x^{-}-y^{-}\right)
B\left({\bf x}_{\perp}-{\bf y}_{\perp}\right)+
b_{20}\left({\bf x}_{\perp}-{\bf y}_{\perp}\right)
\right]
\left(\partial^{-1}_{-}\Psi^{\dagger}_{+}\right)\left(x^{+},y^{-},{\bf y}_{\perp}\right)
\gamma^{-}+
\eeqs
\beqs
\!+\sqrt{2}igM\!
\Big[\!\!
\left(\partial_{-}^{-1}\phi\right)\!\!\left(x^{+},x^{-},{\bf x}_{\perp}\right)\!-\!
\left(\partial_{-}^{-1}\phi\right)\!\!\left(x^{+},y^{-},{\bf y}_{\perp}\right)\!\!
\Big]\!\!
\left[\!
{1\over{4i}}
{\rm{sgn}}\!\left(x^{-}-y^{-}\right)\!
\delta^{(2)}\!\left({\bf x}_{\perp}-{\bf y}_{\perp}\right)\!+\!
a_{0}\!\left({\bf x}_{\perp}-{\bf y}_{\perp}\right)\!
\right]\!\!
\Lambda_{-}+
\eeqs
\beqs
+{i\over\sqrt{2}}g^{2}\!
\Big[\!\!
\left(\partial_{-}^{-1}\phi^{2}\right)\!\!\left(x^{+},x^{-},{\bf x}_{\perp}\right)-
\left(\partial_{-}^{-1}\phi^{2}\right)\!\!\left(x^{+},y^{-},{\bf y}_{\perp}\right)\!\! 
\Big]\!\!
\left[\!
{1\over{4i}}
{\rm{sgn}}\!\left(x^{-}-y^{-}\right)\!
\delta^{(2)}\!\left({\bf x}_{\perp}-{\bf y}_{\perp}\right)\!+\!
a_{0}\!\left({\bf x}_{\perp}-{\bf y}_{\perp}\right)\!
\right]\!\!
\Lambda_{-}.
\eeqs
The equation (\ref{dirbracketphipsimf}), on the other hand, can be expressed as
\beq
\label{dirbracketphipsimfint}
\left[
\Psi_{-}\left(x^{+},x^{-},{\bf x}_{\perp}\right) \; , \; \phi\left(x^{+},y^{-},{\bf y}_{\perp}\right)
\right]=
igf^{\;\rm{reg.}}_{3}\left(x^{+},x^{-},{\bf x}_{\perp},y^{-},{\bf y}_{\perp}\right)+
\eeq
\beqs
+{1\over 2}g\gamma^{+}
\left(\partial^{-2}_{-}\Psi_{+}\right)\left(x^{+},x^{-},{\bf x}_{\perp}\right)
B\left({\bf x}_{\perp}-{\bf y}_{\perp}\right)-
\eeqs
\beqs
-{1\over 2}g\gamma^{+}
\left(\partial^{-1}_{-}\Psi_{+}\right)\left(x^{+},x^{-},{\bf x}_{\perp}\right)
\left[
\left(x^{-}-y^{-}\right)
B\left({\bf x}_{\perp}-{\bf y}_{\perp}\right)+
b_{20}\left({\bf x}_{\perp}-{\bf y}_{\perp}\right)
\right]. 
\eeqs
And then, from the pattern (\ref{dirbracketpartpphipsimf}), we have  
\beq
\label{dirbracketpartpphipsimfint}
\left[
\Psi_{-}\left(x^{+},x^{-},{\bf x}_{\perp}\right) \; , \; \partial_{+}\phi\left(x^{+},y^{-},{\bf y}_{\perp}\right) 
\right]=
igf_{6}^{\;\rm{reg.}}\left(x^{+},y^{-},{\bf y}_{\perp},x^{-},{\bf x}_{\perp}\right)-
\eeq
\beqs
-{i\over{2\alpha}}g\gamma^{+}
\left(\partial^{-2}_{-}\Psi_{+}\right)\left(x^{+},y^{-},{\bf y}_{\perp}\right)
\left[
{1\over{4i}}
\delta^{(2)}\left({\bf x}_{\perp}-{\bf y}_{\perp}\right)
{\rm{sgn}}\left(x^{-}-y^{-}\right)-
a_{0}\left({\bf y}_{\perp}-{\bf x}_{\perp}\right)
\right]+
\eeqs
\beqs
+{1\over 2}g\gamma^{+}
\left(\partial^{-2}_{-}\Psi_{+}\right)\left(x^{+},x^{-},{\bf x}_{\perp}\right)
\left[
{1\over{4\alpha}}
\delta^{(2)}\left({\bf x}_{\perp}-{\bf y}_{\perp}\right)
{\rm{sgn}}\left(x^{-}-y^{-}\right)-
d_{10}\left({\bf y}_{\perp}-{\bf x}_{\perp}\right)
\right]-
\eeqs
\beqs
-{1\over 2}g\gamma^{+}\!
\left(\partial^{-1}_{-}\Psi_{+}\right)\!\!\left(x^{+},x^{-},{\bf x}_{\perp}\right)\!\!
\left[
{1\over{4\alpha}}
\delta^{(2)}\!\left({\bf x}_{\perp}-{\bf y}_{\perp}\right)\!
\mid\!x^{-}-y^{-}\!\mid\!-\!
\left(x^{-}-y^{-}\right)\!
d_{10}\!\left({\bf y}_{\perp}-{\bf x}_{\perp}\right)\!+\!
d_{100}\!\left({\bf y}_{\perp}-{\bf x}_{\perp}\right)
\right]\!. 
\eeqs
The contributions: $f^{\;\rm{reg.}}_{1,3,6}(x^{+},{\bar x},{\bar y})$, $f^{2\;\rm{reg.}}_{1}(x^{+},{\bar x},{\bar y})$ and 
$h^{\rm{reg.}}_{1,2}(x^{+},{\bar x},{\bar y})$ are the regularized limit terms of relevant functions: (\ref{invfinr1elements1}), (\ref{invfinr1elements3}), (\ref{invfinr1elements6}), (\ref{invfinr1elements11}), (\ref{invhfunct1}) and (\ref{invhfunct2}). They require further analysis. 

\section{Limit $\alpha=0$}
\label{limit}

Whenever $\alpha=0$, the Lagrangian density of the light-front Yukawa model without higher order derivatives, discussed from presented point of view in \cite{b20}, is restored. This turn needs to be analyzed a little more specifically, however. The set of the constraints is reduced by one in this case and therefore, the size of the Dirac-Bergmann matrix yields $5\times5$, instead of $6\times6$. The first equations within  (\ref{canhiordscalmom}), (\ref{canquant3}) and also the (\ref{constraintprimary0}) disappear, what causes, that the third row and the third column of the Dirac-Bergmann matrix (\ref{fmatrixelements}), of its decompositions: (\ref{fmatrixfreeelements}), (\ref{gmatrixelements}), (\ref{gtildematrixelement1}), (\ref{gtildematrixelement2}) and of its inverse: (\ref{invfmatrixelements}), (\ref{invfinr1}), (\ref{invfinr2}), (\ref{finvfin}) are eliminated. Thus, the functions $b_{1}(x^{+},{\bar x})$ and $d_{1,2}(x^{+},{\bar x})$ come out from the considerations. This follows, that also $F_{2,4,6,7}(x^{+},{\bar x},{\bar y})$ vanish and only: $F_{1,3,5,8,9}(x^{+},{\bar x},{\bar y})$, 
$F^{2}_{1}(x^{+},{\bar x},{\bar y})$, $H_{1,2,3,4}(x^{+},{\bar x},{\bar y})$ still are present in the Yukawa model with $\alpha=0$. Then, the set of the equations (\ref{acdiffeq}) - (\ref{d2d2diffeq}), being the conditions of the invertibility of the Dirac-Bergmann matrix, is reduced by erasing both the formulae (\ref{b1b1diffeq}) and the first of: (\ref{d1d1diffeq}), (\ref{d2d2diffeq}). The pattern (\ref{b2diffeq}) becomes trivial and the whole set of the studied equations embraces from now both the relationships (\ref{acdiffeq}) and the one $\partial_{-}b_{2}({\bar x})=-(1/2)\delta^{(3)}({\bar x})$, taken from the second expression (\ref{d1d1diffeq}) or (\ref{d2d2diffeq}) in the limit $\alpha=0$. Therefore, the obtained set of equations is the same as in \cite{b20}. We complete this part of the discussion by two comments. The first is, that the function $b_{2}({\bar x})$ changes its character before and after the $\alpha=0$ turn. Before, we have this function described by the (\ref{b2solut}) with $B({\bf x}_{\perp})$ determined by the (\ref{Bsolut}) and after, there is 
$\partial_{-}b_{2}({\bar x})=-(1/2)\delta^{(3)}({\bar x})$, what shows, that the limit $\alpha=0$ is consistent with the general case $\alpha\neq0$, but requires delicate handling. This remains in relation to the second observation, referred to the fact, that the primary system of the equations (\ref{acdiffeq}) - (\ref{d2d2diffeq}) describes both the cases: $\alpha\neq0$ and $\alpha=0$, whereas its consequence, the set (\ref{acdiffeqfin}) - (\ref{b2b2diffeqfin}) with ({\ref{diffsystcond}), is valid only for $\alpha\neq0$ and cannot be applied for taking the aforementioned limit $\alpha=0$. 

\section{Conclusions}
\label{conc}

This work is devoted to the quantization of the light-front Yukawa model in $D=1+3$ dimensions with higher order derivatives of the scalar field, described by the Bernard-Duncan term. The discussion is focused on the problem of obtaining the Dirac brackets and then, the (anti-) commutator algebra of interacting fields in presence of the constraints. The Dirac method for the constraints in the quantum theories and the Ostrogradski formalism for the higher order derivatives are used. The second purpose of this work is to discuss the sensitivity of the results on the modifications, caused by introduction of the aforesaid higher order derivatives, in comparison to the same model without the Bernard-Duncan term. From the mathematical point of view the major concern is to compute the inverse of the functional Dirac-Bergmann matrix with the interactions and with the higher order derivative term. The relevant approach for solving this problem in the case of the array, which includes the differential operators and the fermionic bispinor operators is introduced and applied, what gives the complete algebra of the (anti-) commutators, corresponding to the constraints. Proposed approach is based on the certain identity of the matrix series for the inverted array. This may have two fully equivalent variants. The first, applied in this work, handles only the pure matrix series, whereas the second one emphasizes the array expansion in the powers of the Yukawa coupling constant between fermions and scalar. The latter seems to have more perturbative character. Both, the structure of the light-front Yukawa model with higher order derivatives and the structure of the relevant constraints lead to the finite expansion of the functional inverse of the Dirac-Bergmann matrix. Thus, the first variant of already introduced approach is used here for the calculations. Of course, the second variant of deliberated approach gives the same, truncated results for the concerned (anti-) commutator algebra of interacting fields in the case of our Yukawa model. This contains only the contributions of the first and of the second order of the coupling constant $g$. These results agree with the work \cite{b20}, devoted to the model without higher order derivatives. The next problem arises at this point and demands further analysis. This is dedicated to the general formulation of the functions, which define the free inverse Dirac-Bergmann matrix. 

On the flip side, the structure of the results for the (anti-) commutator algebra is strongly dependent on the higher order derivative  contribution and rapidly changes its character, although finite, in the limit $\alpha=0$, {\it i.e.} wherein the Bernard-Duncan term is switched off. This feature may initialize the searches, interesting from the physical point of view, for such the models, for such the constraints and interactions, which exhibit the infinite character of the series determining aforementioned algebra. It opens the question, for what class of the models the functional inverse Dirac-Bergmann matrix of the constraints with interactions can be computed exactly and for what this can only be derived approximately, giving the infinite, perturbative solution of the established problem. This is a very good subject of the future considerations.\\ 

%\acknowledgement{}

\appendix{{\bf{Appendix A: Heisenberg Equations for Yukawa Model with Higher Order Derivatives}}}\\ 
\label{appa}

The presence of $n$ higher order derivatives of the single scalar field within  
${\cal L}={\cal L}(\phi,\phi_{,\mu_{1}},\dots,\phi_{,\mu_{1}\dots\mu_{n}})$ modifies the density of the canonical energy-momentum tensor. Now, this  object yields in the general case 
\beq
\label{hemtden}
{\cal T}^{\mu\nu}=
\sum_{r=1}^{n}
\left\{
\sum_{s=0}^{n-r}
(-1)^{s}
\left(\prod_{t=1}^{s}\partial_{\rho_{t}}\right)
\left({{\partial{\cal L}}\over{\partial\phi_{,\mu\sigma_{1}\dots\sigma_{r-1}\rho_{1}\dots\rho_{s}}}}\right)
\right\}
\left(\prod_{u=1}^{r-1}\partial_{\sigma_{u}}\right)\partial^{\nu}\phi-
g^{\mu\nu}{\cal L}.  
\eeq
This pattern replaces the standard definition for the ${\cal T}^{\mu\nu}$, which refers to the presence of only the first order derivative of the scalar field: ${\cal T}^{\mu\nu}=(\partial{\cal L}/\partial\phi_{,\mu})\partial^{\nu}\phi-g^{\mu\nu}{\cal L}$. The new formulation of the density of the energy-momentum tensor leads to more severalfold precise structure of the operators $P^{\mu}$ for the translation in the $x^{\mu}$ directions, however their general definition, displayed below, does not change. These operators for the light-front formulation, wherein the variable of evolution is $x^{+}$, obey 
\beq
\label{gentralfmu}
P^{\mu}\left(x^{+}\right)=\!
\int_{R^{3}}\!\!{d^{3}}{\bar y}\;
{\cal T}^{+\mu}\left(x^{+},{\bar y}\right), \;\;\;\;\;\; 
\mu=+,-,j, \;\;\;\;\;\; j=1,2. 
\eeq 
The extension of the Lagrangian density described in this work affects, as a consequence, the results inferred from the Heisenberg equations 
\beq
\label{hempsim}
i\partial_{\mu}\psi_{a}\left(x^{+},{\bar x}\right)=
\left[\psi_{a}\left(x^{+},{\bar x}\right) \; , \; P_{\mu}\left(x^{+}\right)\right], \;\;\;\;\;\;  
\mu=+,-,j, \;\;\;\;\;\;  j=1,2. 
\eeq
In our case is $\psi_{a}=\Psi_{\pm},\Psi^{\dagger}_{\pm},\partial_{\pm}\phi,\partial_{j}\phi$, so the Heisenberg equations relate to the fields and to the derivatives of the field, which are present in this considerations. These equations \cite{b28} may directly allow to deduce some elements of the (anti-) commutator algebra for discussed model, but without appealing to the canonical quantization rules or other methods \cite{b29,b30,b31}, being to our disposal. We obtain, using the Lagrangian density (\ref{ldymlf}) and the prescription (\ref{hemtden}), that the ${\cal T}^{++}$ component of the density of the canonical energy-momentum tensor is 
\beq
\label{emtpp}
{\cal T}^{++}=
i\sqrt{2}\Psi^{\dagger}_{+}\partial_{-}\Psi_{+}+
\left(\partial_{-}\phi\right)^{2}-
\alpha\left(\partial_{-}\partial^{2}\phi\right)
\left(\partial_{-}\phi\right)+
\alpha\left(\partial^{2}\phi\right)
\left(\partial^{2}_{-}\phi\right).
\eeq 
It is not difficult to notice, that by the same method we can compute 
\beq
\label{emtpmd}
{\cal T}^{+-}=
\eeq
\beqs
=-i\sqrt{2}\;
\Psi^{\dagger}_{-}\partial_{-}\Psi_{-}-
{i\over{\sqrt{2}}}\Psi^{\dagger}_{+}\gamma^{-}
\gamma^{j}\partial_{j}\Psi_{-}-
{i\over{\sqrt{2}}}\Psi^{\dagger}_{-}\gamma^{+}
\gamma^{j}\partial_{j}\Psi_{+}+
{1\over{\sqrt{2}}}M
\Psi^{\dagger}_{+}\gamma^{-}\Psi_{-}+
{1\over{\sqrt{2}}}M
\Psi^{\dagger}_{-}\gamma^{+}\Psi_{+}+
\eeqs
\beqs
+{1\over 2}
\left(\partial_{j}\phi\right)
\left(\partial_{j}\phi\right)+
{1\over 2}m^{2}\phi^{2}+
{1\over 4}\lambda\phi^{4}+
{1\over{\sqrt{2}}}g
\phi\;\Psi^{\dagger}_{+}\gamma^{-}\Psi_{-}+
{1\over{\sqrt{2}}}g
\phi\;\Psi^{\dagger}_{-}\gamma^{+}\Psi_{+}-
\alpha\left(\partial_{-}\partial^{2}\phi\right)
\left(\partial_{+}\phi\right)+ 
{1\over 2}
\alpha\left(\partial^{2}\phi\right)
\left(\Delta_{\perp}\phi\right). 
\eeqs
Some of the terms in this formula may be substituted with help of the first equations (\ref{empsim}) and (\ref{empsim1}), what reduces the final expression for the ${\cal T}^{+-}$ component to 
\beq
\label{emtpmk} 
{\cal T}^{+-}=
i\sqrt{2}\;
\Psi^{\dagger}_{+}\partial_{+}\Psi_{+}+
{1\over 2}
\left(\partial_{j}\phi\right)
\left(\partial_{j}\phi\right)+
{1\over 2}m^{2}\phi^{2}+
{1\over 4}\lambda\phi^{4}-
\alpha\left(\partial_{-}\partial^{2}\phi\right)
\left(\partial_{+}\phi\right)+ 
{1\over 2}\alpha\left(\partial^{2}\phi\right)
\left(\Delta_{\perp}\phi\right). 
\eeq
On the other hand, the ${\cal T}^{+j}$ satisfies 
\beq
\label{emtpj}
{\cal T}^{+j}=
-i\sqrt{2}\;
\Psi^{\dagger}_{+}\partial_{j}\Psi_{+}-
\left(\partial_{-}\phi\right)
\left(\partial_{j}\phi\right)+
\alpha\left(\partial_{-}\partial^{2}\phi\right)
\left(\partial_{j}\phi\right)-
\alpha\left(\partial^{2}\phi\right)
\left(\partial_{-}\partial_{j}\phi\right).
\eeq

Now, we take the Heisenberg equations (\ref{hempsim}) to obtain possible (anti-) commutators for the Yukawa model with higher order derivatives and briefly report the results. This approach gives the same relationships as: (\ref{dirbracketpsippsipf}), (\ref{dirbracketpsippsidpf}), (\ref{dirbracketpsimpsipf}), (\ref{dirbracketpsimpsidpf}), (\ref{dirbracketphipsipf}), (\ref{dirbracketpsippartpphi}), (\ref{dirbracketphiparpphif}) or these, which are the corollaries of the itemized. The method based on the Heisenberg equations does not give any predictions for the rules (\ref{dirbracketpsimpsimf}) and (\ref{dirbracketpsimpsidmf}). The patterns: (\ref{dirbracketphipsimf}), (\ref{dirbracketpartpphipsimf}) and (\ref{dirbracketparpphiparpphif}) have the analogues derived from the aforementioned approach, but not consistent with them. They are trivial: $[\Psi_{-}(x^{+},{\bar x}),\partial_{-}\phi(x^{+},{\bar y})]=0$, 
$[\Psi_{-}(x^{+},{\bar x}),\partial_{+}\phi(x^{+},{\bar y})]=0$, $[\partial_{+}\phi(x^{+},{\bar x}),\partial_{+}\phi(x^{+},{\bar y})]=0$. The analogue of the result (\ref{dirbracketphiphif}), computed from the Heisenberg equations, splits, due to the fact, that in this approach all the derivatives of the scalar field are independent, whereas for our main considerations, based on the Dirac procedure, the independent derivative is only $\partial_{+}\phi$. And that's why we obtained from the Heisenberg equations the following commutator expressions: 
$[\phi(x^{+},{\bar x}),\phi(x^{+},{\bar y})]=0$, $[\phi(x^{+},{\bar x}),\partial_{j}\phi(x^{+},{\bar y})]=0$, 
$[\phi(x^{+},{\bar x}),\partial_{-}\phi(x^{+},{\bar y})]=(i/2)\delta^{(3)}({\bar x}-{\bar y})$. Remarkably, only the last one matches the (\ref{dirbracketphiphif}) and is valid in the limit $\alpha=0$. For the same reason we put here the relationship 
$\{\Psi_{+}(x^{+},{\bar x}),\partial_{+}\Psi_{+}(x^{+},{\bar y})\}=0$, but now non-conflicting with the (\ref{dirbracketpsippsipf}). 

To sum up, the method based on the Heisenberg equations is not sensitive to the structure of the constraints and to the interactions for the Yukawa model with higher order derivatives. This method is not constructional, whereas the Dirac approach is. It is therefore only the last one gives the systematic approach for the quantization in the presence of the constraints with interactions.\\ 

\appendix{{\bf{Appendix B: Light-Front Formulation}}}\\
\label{appb}

The light-front coordinates for the flat space-time of $D=1+3$ dimensions obey: 
\beq
\label{lfcoord}
x=
\left(x^{+},x^{-},{\bf x}_{\perp}\right), \;\;\;\;\;\; 
x^{\pm}=
{1\over{\sqrt{2}}}\left(x^{0}\pm{x^{3}}\right), \;\;\;\;\;\;  
{\bf x}_{\perp}=
\left(x^{1},x^{2}\right). 
\eeq
The hyper-surface of the quantization $x^{+}=0$ for the light-front and respectively, the $x^{-}=0$ for the anti-light-front, enable us to introduce the reduced coordinates: 
\beq
\label{lfhypcoord}
{\bar x}=
\left(x^{-},{\bf x}_{\perp}\right)\simeq
\left(x^{+}=0,x^{-},{\bf x}_{\perp}\right), \;\;\;\;\;\;  
{\underline x}=
\left(x^{+},{\bf x}_{\perp}\right)\simeq
\left(x^{+},x^{-}=0,{\bf x}_{\perp}\right).
\eeq
The components of the metric tensor, relevant to the coordinates (\ref{lfcoord}), are:  
\beq
\label{gtensorcomp}
g_{+-}=1, \;\;\;\;\;\;
g_{++}=0=g_{--}, \;\;\;\;\;\; 
g_{{\pm}j}=0, \;\;\;\;\;\; 
g_{jk}=-\delta_{jk}, \;\;\;\;\;\; 
j,k=1,2. 
\eeq
They allow to lift up and to pull down the indexes in following way: 
\beq
\label{lifting}
a_{\pm}=a^{\mp}, \;\;\;\;\;\; 
a_{j}=-a^{j}, \;\;\;\;\;\; 
j=1,2. 
\eeq
Therefore, the scalar product of the four-vectors satisfies: 
\beq
\label{scalprod}
a\cdot{b}=
a_{+}b_{-}+a_{-}b_{+}-
{\bf a}_{\perp}\cdot{\bf b}_{\perp}, \;\;\;\;\;\;  
{\bf a}_{\perp}\cdot{\bf b}_{\perp}=
a_{j}b_{j}, \;\;\;\;\;\; 
j=1,2. 
\eeq
Herein, the Einstein notation is introduced. Accordingly, the Dirac slash in the light-front formulation yields: 
\beq
\label{slash}
\not\!a=
a_{\mu}\gamma^{\mu}=
a_{+}\gamma^{+}+a_{-}\gamma^{-}-
\not\!{\bf a}_{\perp}, \;\;\;\;\;\; 
\not\!{\bf a}_{\perp}=
a_{j}\gamma^{j}, \;\;\;\;\;\;
\mu=+,-,j, \;\;\;\;\;\; 
j=1,2.
\eeq
The coordinates (\ref{lfcoord}) lead to the light-front derivatives: 
\beq
\label{derivlf}
\partial_{\pm}=
{{\partial}\over{\partial{x^{\pm}}}}, \;\;\;\;\;\; 
\partial_{j}=
{{\partial}\over{\partial{x^{j}}}}, \;\;\;\;\;\; 
j=1,2  
\eeq
and to the Laplace or to the d'Alembert operators:  
\beq
\label{dal}
\Delta_{\perp}=
\partial_{j}\partial_{j}, \;\;\;\;\;\;   
\partial^{2}\equiv\Box=
\partial_{+}\partial_{-}+
\partial_{-}\partial_{+}-
\Delta_{\perp}, \;\;\;\;\;\; 
j=1,2, 
\eeq
wherein the last one is in the symmetrized form, required for the formalism presented in this work. The definition of the light-front Dirac gamma matrices is relevant to the (\ref{lfcoord}). Thus: 
\beq
\label{defgammapm}
\gamma=
\left(\gamma^{+},\gamma^{-},\gamma_{\perp}\right), \;\;\;\;\;\; 
\gamma^{\pm}=
{1\over{\sqrt{2}}}
\left(\gamma^{0}\pm{\gamma^{3}}\right), \;\;\;\;\;\;  
\gamma_{\perp}=
\left(\gamma^{1},\gamma^{2}\right). 
\eeq
There are some characteristic properties of these arrays, used in this considerations: 
\beq
\label{gammamatrixprop}
\left(\gamma^{\pm}\right)^{2}=0, \;\;\;\;\;\;  
\left(\gamma^{\pm}\right)^{\dagger}=\gamma^{\mp}, \;\;\;\;\;\; 
\left(\gamma^{j}\right)^{\dagger}=-\gamma^{j}, \;\;\;\;\;\; 
j=1,2. 
\eeq
There is the natural decomposition of the fermionic bispinor 
\beq
\label{spindecomppsi}
\Psi=
\Psi_{+}+\Psi_{-},   
\eeq
whereas the spinors $\psi_{\pm}$ arrange the components of the above pattern:
\beq
\label{spinors}
\Psi=
\left(
\begin{array}{c}
\psi_{+}\\
\psi_{-}\\
\end{array}
\right), \;\;\;\;\;\; 
\Psi_{+}=
\left(
\begin{array}{c}
\psi_{+}\\
0\\
\end{array}
\right)
\simeq\psi_{+}, \;\;\;\;\;\; 
\Psi_{-}=
\left(
\begin{array}{c}
0\\
\psi_{-}\\
\end{array}
\right)
\simeq\psi_{-}. 
\eeq
The two constant bispinors $u_{\pm}$ allow to perform the inverse decomposition: 
\beq
\label{uconstpmspin}
u_{+}=
\left(
\begin{array}{c}
1\\
0\\
\end{array}
\right), \;\;\;\;\;\;
u_{-}=
\left( 
\begin{array}{c}
0\\
1\\
\end{array}
\right), \;\;\;\;\;\; 
\psi_{\pm}=
u^{\dagger}_{\pm}\Psi.  
\eeq
Their tensor product defines the matrices $\Lambda_{\pm}$, due to:  
\beq
\label{tenprodlambdapm}  
u_{\pm}\otimes{u^{\dagger}_{\pm}}=
\Lambda_{\pm}=
u^{\dagger}_{\pm}\otimes{u_{\pm}}, \;\;\;\;\;\; 
\Lambda_{+}=
\left(
\begin{array}{cc} 
1&\;\;0\\
0&\;\;0\\
\end{array}
\right), \;\;\;\;\;\; 
\Lambda_{-}=
\left(
\begin{array}{cc} 
0&\;\;0\\
0&\;\;1\\
\end{array}
\right).  
\eeq
These $\Lambda_{\pm}$ arrays permit us to write the decomposition (\ref{spindecomppsi}) as:  
\beq
\label{psipmaslambdapsi}
\Psi_{+}=\Lambda_{+}\Psi, \;\;\;\;\;\;  
\Psi_{-}=\Lambda_{-}\Psi.
\eeq
The aforementioned matrices are projective and Hermitian: 
\beq
\label{projprop}
\Lambda_{\pm}^{2}=
\Lambda_{\pm}, \;\;\;\;\;\; 
\Lambda_{\pm}\Lambda_{\mp}=
0=
\Lambda_{\mp}\Lambda_{\pm}, \;\;\;\;\;\; 
\Lambda_{+}+\Lambda_{-}={\rm I}, \;\;\;\;\;\; 
\Lambda_{\pm}^{\dagger}=
\Lambda_{\pm}.
\eeq
We can easily provide the $\Lambda_{\pm}$ matrices by the Dirac gamma ones 
\beq
\label{lambdabygamma}
\Lambda_{\pm}=
{1\over 2}
\gamma^{\mp}\gamma^{\pm}.
 \eeq
The spinor decomposition, applied here, enables us to put the $\gamma^{\pm}$ arrays as the following tensor products: 
\beq
\label{gammabyu}
u_{\mp}\otimes{u^{\dagger}_{\pm}}=
\gamma^{\pm}=
u^{\dagger}_{\pm}\otimes{u_{\mp}}, \;\;\;\;\;\; 
\gamma^{+}=
\left(
\begin{array}{cc} 
0&\;\;0\\
1&\;\;0\\
\end{array}
\right), \;\;\;\;\;\; 
\gamma^{-}=
\left(
\begin{array}{cc} 
0&\;\;1\\
0&\;\;0\\
\end{array}
\right). 
\eeq
There are some handling rules of the $\gamma^{\pm}$ and the $\Lambda_{\pm}$ matrices. They satisfy 
\beq
\label{gammalambdainterm}
\gamma^{\pm}\Lambda_{\pm}=
\gamma^{\pm}=
\Lambda_{\mp}\gamma^{\pm}.
\eeq
The $\gamma^{\pm}$ arrays allow to do certain kinds of the projections of the $\Psi$ and the $\Psi_{\mp}$ bispinors with the flips of their internal structure:
\beq
\label{gammapsi}
\gamma^{+}\Psi=
\left(
\begin{array}{c}
0\\
\psi_{+}\\
\end{array}
\right), \;\;\;\;\;\;
\gamma^{-}\Psi=
\left(
\begin{array}{c}
\psi_{-}\\
0\\
\end{array}
\right), \;\;\;\;\;\;
\gamma^{\pm}\Psi_{\mp}=0.
\eeq
All the computations within this work are based on the set of presented here formulae and definitions.

\end{document}